\documentclass[12pt,letterpaper]{article}
	\pdfoutput=1
	\usepackage[colorlinks=true,
	linkcolor=Red, 
	citecolor=Red,
	filecolor=Red,
	urlcolor=Red,
	linktoc=page, %%%
	pdfstartview=FitV,
	bookmarksopen=true]{hyperref}
	\usepackage[left=2cm,top=1cm,right=3cm,nohead]{geometry}
	\usepackage[dvipsnames]{xcolor}
	\usepackage{colortbl,cancel}
	\usepackage[utf8]{inputenc} 
	\usepackage{color}
	\usepackage{epsfig,latexsym,amsfonts,mathtools,amsthm,amssymb,amsbsy,multirow,slashed,
		mathrsfs,wasysym,textcomp,subfigure,wrapfig,comment,bbold,array,longtable,multirow,setspace,amsmath,pifont,nicefrac}
	\usepackage{tabularx}
	\usepackage[]{mdframed}
	\usepackage{graphicx}
	\usepackage{setspace}
	\usepackage{subfigure}
	\usepackage{multirow}
	\usepackage[bf]{caption}
	\usepackage{braket}
	\usepackage{comment}
	\usepackage{float}
	\usepackage{tikz}
	\usetikzlibrary{mindmap,trees,shadows}
	\colorlet{color1}{gray!25}
	\usetikzlibrary{positioning}
	\usetikzlibrary{intersections} 
	\newlength{\PicScale}
	\usepackage[UKenglish]{babel}
	\usepackage[toc,page]{appendix}
	\usepackage{hhline}
	\usepackage{geometry}
	\usepackage{cite}
	\usepackage{datetime}
	\usepackage{bbm} 
	\usepackage{bm} 
	\hypersetup{
		pdftitle={},
		pdfauthor={},
		pdfsubject={}
	}

	\newcolumntype{M}[1]{>{\centering\arraybackslash}m{#1}}
	\newcolumntype{N}{@{}m{0pt}@{}}

	\numberwithin{equation}{section}

	\makeatletter
	\def\@cline#1-#2\@nil{%%%
		\omit
		\@multicnt#1%%%
		\advance\@multispan\m@ne
		\ifnum\@multicnt=\@ne\@firstofone{&\omit}\fi
		\@multicnt#2%%%
		\advance\@multicnt-#1%%%
		\advance\@multispan\@ne
		\leaders\hrule\@height\arrayrulewidth\hfill
		\cr
		\noalign{\nobreak\vskip-\arrayrulewidth}}
	\makeatother
\begin{document}
		\pagestyle{empty}
		\begin{center}        % Main title
			{\bf\LARGE T-duality for non-critical heterotic strings\\ [3mm]}
			%\vskip 0.1cm

			\large{H\'ector Parra De Freitas
				\\[2mm]}

			{\small Jefferson Physical Laboratory, Harvard University\\ [-1mm]}
			{\small\textit{Cambridge, MA 02138, USA}\\[0.2cm]}
			
			{\small \verb" hparradefreitas@fas.harvard.edu"\\[-3mm]}
			\vspace{0.3in}
	
			\small{\bf Abstract} \\[3mm]\end{center}
		We consider non-critical heterotic strings compactified on $S^1$. For full rank theories, they are related to odd self-dual lattices and are structurally of the same form as the critical non-supersymmetric theories. For dimensions up to 14 the associated moduli spaces are Coxeter polytopes already studied by Vinberg and Kaplinskaya. In the heterotic string context, the Coxeter diagrams of these moduli spaces are related through transformations representing the process of dimension changing tachyon condensation of Hellerman-Swanson. For dimensions 8 and 6 respectively on $S^1$ and $T^2$ we show that at special points in the moduli space the subcritical string is the CHS background for two coincident NS5-branes and the intersection of two such pairs. These configurations are interpreted as an end result of condensing heterotic winding tachyons along one or two Scherck-Schwarz circles at self-dual radius. We give evidence that in the first case there is a T-duality between the pair of NS5-branes and a recently constructed non-supersymmetric heterotic 6-brane.

	\newpage

	%\tableofcontents

	%%%%%%%%%%%                 %%%%%%%%%%%%%%%%%%%
	%%%%%%%%%%%  DOCUMENT BODY  %%%%%%%%%%%%%%%%%%%
	%%%%%%%%%%%                 %%%%%%%%%%%%%%%%%%%
	
	\setcounter{page}{1}
	\pagestyle{plain}
	\renewcommand{\thefootnote}{\arabic{footnote}}
	\setcounter{footnote}{0}
	%----------------------------------------------------------------------%
	%  Paper begins
	%----------------------------------------------------------------------%
	
	\tableofcontents	
	\newpage

\section{Introduction}
It is a classical result in string theory (see e.g. \cite{Blumenhagen:2013fgp}) that heterotic superstrings compactified on a circle admit marginal deformations spanning the coset
\begin{equation}\label{Msusy}
	\mathcal{M} \simeq O(\text{II}_{1,17})\backslash O(1,17;\mathbb{R})/O(17;\mathbb{R})\,,
\end{equation}
where $O(\text{II}_{1,17})$ is the group of automorphisms of the even self-dual lattice $\text{II}_{1,17}$. These deformations correspond to varying vacuum expectation values of 17 target space moduli fields, namely the circle radius and 16 constant gauge fields (Wilson lines) along the circle in the Cartan subgroup of the gauge group $E_8\times E_8$ (HE) or $Spin(32)/\mathbb{Z}_2$ (HO) \cite{Narain:1985jj,Narain:1986am}.

On physical grounds we expect that at tree level this situation extends to any heterotic worldsheet with a periodic coordinate field, say $X_9$, and target space gauge bundle. One expects to have $1+n$ independent marginal deformations, where $n$ is the rank of the gauge group, giving rise to a moduli space locally described by $O(1,1+n;\mathbb{R})/O(1+n;\mathbb{R})$, and it remains to determine the arithmetic subgroup describing its global structure.\footnote{ When thinking about the global structure of moduli spaces it helps to refer to the simpler example of complex structures of elliptic curves, where the relevant space is the quotient of the upper half plane $\mathbb{H}^+$ by the group of Möbius transformations $SL(2,\mathbb{Z})$. There are two finite distance singularities at $\tau = i$ and $\tau = e^{2\pi i/3}$ as well as an infinite distance singularity at the limit $\tau \to i \infty$, which can be associated to two non-Abelian symmetry enhancements and a decompactification limit in $T^2$ compactifications of bosonic/heterotic strings \cite{Blumenhagen:2013fgp}. Likewise, the moduli space $\mathcal{M}$ in \eqref{Msusy} has a 44 finite distance singularities and 2 infinite distance singularities, which physically correspond to vacua with maximal nonabelian gauge symmetry as well as the decompactification limits to HE and HO \cite{Cachazo:2000ey,Fraiman:2018ebo,Font:2020rsk}.} 
 Often times, this group is given by automorphisms of a certain hyperbolic lattice $\Gamma_{1,1+n}$, and in many situations it can be described in a controlled manner, see e.g. \cite{Vinberg:1972}.
 
In this paper we will consider circle compactifications of heterotic strings of non-critical type \cite{Chamseddine:1991qu}. These theories have $D \neq 10$ target spacetime dimensions as well as a linear dilaton profile necessary e.g. to avoid the conformal anomaly. Supercritical strings ($D > 10$) have a time-like linear dilaton and yield time-dependent backgrounds;\footnote{The physical relevance of these supercritical theories is not too clear, but we are interested in their formal properties insofar as they could be used to understand those of critical theories.} they have been studied in detail in \cite{Hellerman:2004zm}. Subcritical strings ($D < 10$) on the other hand exhibit a spacelike linear dilaton, and may be related to near-horizon backgrounds of various brane configurations, see e.g. \cite{Murthy:2006eg}. As a first result we will show in Section \ref{s:noncritS1} that circle compactification results in a classical moduli space
\begin{equation}
	\mathcal{M}_{1,1+n} \simeq O(1,1+n;\mathbb{Z})\backslash O(1,1+n;\mathbb{R})/O(1+n;\mathbb{R})\,,
\end{equation}
where $O(1,1+n;\mathbb{Z})$ is the group of automorphisms of the odd self-dual lattice $\text{I}_{1,1+n}$ or equivalently the group of integral matrices preserving the quadratic form $-x_0^2+x_1^2+\cdots x_n^2$. We will then transpose known results in the theory of hyperbolic reflection groups \cite{Vinberg:1972,Kaplinskaya:1978} to describe these spaces diagramatically and extract physical data. In the special case of $D = 2$, these results will complement previous investigations of the moduli space carried out in \cite{Seiberg:2005nk,Davis:2005qe,Davis:2005qi} where the circle is thermal (timelike).

Subcritical strings have also received renewed attention as they seem to describe the linear dilaton background of non-supersymmetric heterotic branes whenever the theory is tachyon-free \cite{Kaidi:2023tqo}. There are four such theories, with gauge groups $G = E_8, SU(16)/\mathbb{Z}_2, E_7\times E_7, Spin(24)/\mathbb{Z}_2$, respectively in $D = 9,8,6,2$; there are corresponding non-supersymmetric $p$-branes with $p = 7,6,4,0$. The cases $p = 0,4$ are known from \cite{Polchinski:2005bg,Bergshoeff:2006bs}, while the cases $p = 6,7$ are given in \cite{Kaidi:2023tqo}. Moreover, these theories can arise as the end-point of a tachyon condensation process in a tachyonic $D = 10$ non-supersymmetric string \cite{Kaidi:2020jla}. In Section \ref{s:branes} we consider how this picture is enriched when there is a compact direction (we do not consider the special case $p = 7$), and find among other things that
\begin{itemize}
	\item The linear dilaton background of the 6-brane compactified on a circle with suitable Wilson lines becomes supersymmetric, and is T-dual to the background of two coincident NS5-branes given by the CHS model \cite{Callan:1991at}.
	
	\item In turn, the two NS5-branes can be thought of as the end-point of tachyon condensation for the two winding tachyons of a Scherk-Schwarz reduction of HO at self-dual radius. This result generalizes to any heterotic string with or without supersymmetry.
\end{itemize} 
Motivated by this physical picture we will intertwine our discussion of the moduli spaces $\mathcal{M}_{1,1+n}$ and symmetry enhancements with considerations of tachyon condensation and interpret them, at least formally, to be connected through this process. 

Before proceeding, we make a note that since the backgrounds that we consider in this paper are not to be interpreted as genuine lower dimensional quantum gravity vacua, it is not clear a priori what lessons we can draw from them in regards to the structure of the string landscape. It is the case however that they are obtained through tachyon condensation in critical strings which are known to interpolate with supersymmetric strings. Their study may then shed light into the nature of tachyons in string theory. More precisely, we would like to understand why tachyons appear in a generic way when supersymmetry is broken (see \cite{Baykara:2024tjr,Angelantonj:2024jtu} for recent constructions of heterotic models without tachyons in various dimensions). Rather than take this to be an illness of the theory, we take the point of view that their appearance should be systematically understood, in this case restricting our attention to heterotic strings. In this paper we show in particular that there is a relationship between tachyonic states appearing in a Scherck-Schwarz reduction of any heterotic string are related to the existence of NS5 branes in the original supersymmetric theory.

This paper is ornanized as follows. In Section \ref{s:noncritS1} we work out the global structure of circle compactifications of non-critical heterotic strings and how this information is represented using Coxeter diagrams. In Section \ref{s:maximal} we show how every point of maximal symmetry enhancement can be obtained from these Coxeter diagrams, or alternatively from a higher dimensional tachyonic theory through tachyon condensation. In Section \ref{s:branes} we interpret our results in terms of heterotic branes and derive various relationships among them. We provide our conclusions in Section \ref{s:conc}.

\section{Non-critical heterotic strings on $S^1$}\label{s:noncritS1}
In this section we briefly review the construction of non-critical $Spin(2n)$ heterotic strings and compactify them on a circle $S^1$. We then determine their T-duality group to be the group of automorphisms of the lattice $\Gamma_{1,1}\oplus D_n$, equivalent to $O(1,1+n;\mathbb{Z})$. This group determines the global structure of the moduli space.\footnote{By moduli space we mean the space of marginal deformations of the heterotic worldsheet. In general the usual moduli fields are massless up to a shift in $m^2$, see below.} Its reflective part is seen to be encoded in a Coxeter diagram, from which one easily reads in particular all infinite distance limits comprising every rank $n$ non-critical heterotic string in the corresponding number of spacetime dimensions. We then comment on how these diagrams help to understand the effect of tachyon condensation in the compactified theories. Particular attention is given to the theories with $13 \leq n\leq 18$ i.e. $4 \leq D \leq 14$ with $D$ the number dimensions before compactification. The special case $D = 2$ is considered at the end.
\subsection{$Spin(2n)$ heterotic strings}
Non-critical strings can be studied starting from the simplest family of heterotic strings, which in the fermionic formulation consists of theories with $2n$ free left-moving Majorana-Weyl (MW) fermions $\lambda^A$, $A = 1,...,2n$, in the internal field theory.\footnote{We choose this frame since, unlike other heterotic strings, this family has a representative for every allowed number of target space dimensions. T-duality with the remaining heterotic strings then makes the analysis in this frame completely general. } For $n = 16$ the central charge contribution is $c_L^{\text{int}} = 16$ and the theory is critical with ten spacetime dimensions in target space. The 32 fermions give rise to an $\mathfrak{so}_{32}$ gauge algebra, but working out the spectrum one finds states transforming as vectors, spinors and co-spinors of $Spin(32)$ hence the full gauge group is simply connected. We find in particular a spacetime tachyon transforming in the vector representation. This is one of the tachyonic heterotic strings discovered in \cite{Kawai:1986vd,Seiberg:1986by}. 

For generic $n$, we have $c_L^{\text{int}} = n$, and one must introduce a linear dilaton background to avoid a conformal anomaly. Overall consistency requires changing the number of spacetime dimensions. The right-moving part of the heterotic worldsheet has $N = 1$ supersymmetry, hence for each right-moving coordinate field $X_R^\mu$ there is a corresponding superpartner fermion $\psi_R^\mu$. Since a linear dilaton background contributes both to $c_L$ and $c_R$ by the same amount, changing the number of $\lambda$ fields by $\Delta n$ requires doing the same for $X_L, X_R$ and $\psi_R$. Taking $\Phi = -V_\mu X^\mu$, we have
\begin{equation}
	c_\Phi = 6\alpha' V_\mu V^\mu\,,
\end{equation}
and so we require $6 \alpha' V_\mu V^\mu = -\Delta n$. From now on we will set $\alpha' = 1$. For positive $n$ we obtain supercritical strings with a time-like linear dilaton, and for negative $n$ we obtain subcritical strings with a space-like linear dilaton. 

Working in lightcone gauge, the torus partition function $\mathcal{Z}(\tau,\bar \tau)$ can be written in the bosonic formulation in terms of conjugacy classes of $Spin(2n)$; the linear dilaton does not enter explicitly, although its presence is felt by the effective change in the number spacetime dimensions $D$. This function is of the form\footnote{c.f. eq. A.8 in \cite{Berasaluce-Gonzalez:2013sna} with their $n$ our $2(n-16)$.} 
\begin{equation}\label{ZD}
\mathcal{Z}^D(\tau,\bar \tau) = \frac{1}{(\sqrt \tau_2 \eta \bar \eta)^{D-2}}\left(O_{2n}\bar V_{D-2} + V_{2n} \bar O_{D-2} - S_{2n} \bar S_{D-2} - C_{2n} \bar C_{D-2} \right)\,,	
\end{equation}
where $\eta$ is the Dedekind eta function and every function except $\tau_2 = \text{Im}\tau$ in the RHS is holomorphic (unbarred) or anti-holomorphic (barred). We use the $Spin(2n)$ characters 
\begin{equation}
	\begin{split}
		O_{2n} &= \frac{1}{2\eta^{n}}(\vartheta_3^n + \vartheta_4^n)\,, ~~~~~ V_{2n} = \frac{1}{2\eta^n} (\vartheta_3^n - \vartheta_4^n)\,,\\
		S_{2n} &= \frac{1}{2\eta^n} (\vartheta_2^n + \vartheta_1^n)\,, ~~~~~ C_{2n} = \frac{1}{2\eta^n} (\vartheta_2^n - \vartheta_1^n)\,,
	\end{split}
\end{equation}
with $\vartheta_{1,2,3,4}$ the usual Jacobi theta functions evaluated at zero chemical potential. From the analysis above, there is a constraint $D = 10 + 2n - 32$, and so $2n \geq 24$ with the lower bound saturated by the subcritical 2D heterotic string with gauge group $Spin(24)$. The supercritical case is referred to as $\text{HO}^{+(m)/}$ in its first extensive study \cite{Hellerman:2004zm},  where $m$ is the number of dimensions beyond 10, i.e. $m = 2(n-16)$. 

\subsection{Circle compactification}
Compactifying our $Spin(2n)$ theory on a circle is done by exchanging
\begin{equation}
	\frac{1}{\sqrt{\tau_2} \eta \bar \eta} \to \frac{1}{\eta \bar \eta}\sum_{(p_L,p_R)\in \Gamma_{1,1}} q^{\tfrac12 p_L^2}\bar q^{\tfrac12 p_R^2}\,,
\end{equation}
with $\Gamma_{1,1}$ the unique even self-dual lattice of signature (1,1). This factor can be combined with each of the conjugacy classes inside the parentheses in \eqref{ZD}, and it is convenient to write the overall partition function of the compactified theory as
\begin{equation}\label{ZDS1}
	\mathcal{Z}^D_{S^1}(\tau,\bar \tau) = \frac{1}{(\sqrt \tau_2 \eta \bar \eta)^{D-3}}\left(\mathcal{Z}_{v,n}\bar V_{D-2} + \mathcal{Z}_{o,n} \bar O_{D-2} - \mathcal{Z}_{s,n} \bar S_{D-2} - \mathcal{Z}_{c,n} \bar C_{D-2} \right)\,,	
\end{equation}
with
\begin{equation}\label{Zomegan}
	\mathcal{Z}_{\omega,n} = \frac{1}{\eta^{n+1}\bar \eta} \sum_{(P_L,p_R) \in \Gamma^\omega_{1,n+1}} q^{\tfrac12 P_L^2}\bar q^{\tfrac12 p_R^2}\,, ~~~~~ \omega = v,o,s,c\,.
\end{equation}
Here $\Gamma^v_{1,n+1}$ is the hyperbolic lattice $\Gamma_{1,1}\oplus D_n$, and the remaining sets correspond to replacing $D_n$ by $D_n + y$ with $y$ a non-trivial element in the set of conjugacy classes $D_n^*/D_n$. This light-cone computation cannot be performed for the special case $D = 2$; we will come back to this problem in Section \ref{ss:2dstrings}.

We split $P_L = (P,p_L)$ with $P, p_L$ the gauge lattice and circle contributions respectively, so that the momenta take the form
\begin{align}
	P 	&= \pi + Aw\,,\label{momentum-P} \\ 
	p_L &= \frac{1}{\sqrt{2}R}\left(n + \left(R^2 -\frac12 A^2\right)w  - A \cdot \pi\right)\,,\label{momentum-pL}\\ 
	p_R &= \frac{1}{\sqrt{2}R}\left(n - \left(R^2 +\frac12 A^2\right)w  - A \cdot \pi\right)\,.\label{momentum-pR} 
\end{align}
$R$ is the circle radius and we have turned on arbitrary Wilson line moduli $A = (A_1,...,A_{n})$ for the gauge group $Spin(2n)$ along $S^1$; $n \in \mathbb{Z}$ denotes the Kaluza-Klein momentum, $w \in \mathbb{Z}$ the winding number and $\pi \in D_n+y$ the momentum along the internal $D_n$ lattice or its other conjugacy classes depending on the sector under consideration. 

The spectrum of the theory suffers an effective mass shift on right-moving NS states due to the presence of the linear dilaton, reflected in the change of characters in the partition function relative to the critical case. The mass formula and level matching conditions thus read
\begin{align}
	\frac{1}{2}m^2 &= \frac{1}{2}P_L^2 + \frac12 p_R^2 + N_L + N_R - \frac{1}{2}\delta_D-
	\begin{cases}
		1 & \text{R sector}\\
		\frac32  & \text{NS sector}
	\end{cases} \,,\label{massform}\\ 
	0 &= \frac{1}{2}P_L^2 - \frac{1}{2}p_R^2 + N_L - N_R - 
	\begin{cases}
		1 & \text{R sector}\\
		\frac12  & \text{NS sector}
	\end{cases}\label{levelmatch}\,,
\end{align}
where $N_{L,R}$ are oscillator numbers and $\delta_D\equiv (10-D)/8$ is the aforementioned shift in $m^2$. For massless states in the $R$ sector, this shift is canceled by an equivalent shift in $P_L^2$ resulting from the reduction of the gauge bundle. We will come back to this point in Section \ref{s:maximal}.

We will usually refer to states in the NS sector as being massless when their effective mass is due solely to the linear dilaton, i.e. when $m^2 = (10-D)/8$, and similarly for tachyons. Up to this shift, the theory generically has the usual graviton, B-field and dilaton as well as 18 gauge bosons furnishing the gauge group $U(1)^{17}_L \times U(1)_R$, and no massless fermions. At special points in the moduli space, however, we expect to have extra low lying states in each of the four conjugacy classes, including massless gauge bosons, tachyons as well as massless spinors and conjugate spinors. The problem of determining these enhancements was addressed in detail in \cite{Fraiman:2023cpa} for the critical case, and we will see here that for non-critical strings the situation is analogous.

\subsection{The moduli space}  
At the local level, the moduli space of our circle-compactified theory is the $(n+1)$-dimensional hyperbolic space
\begin{equation}
	\mathcal{M}_{1,n+1} = O(1,1+n)/O(1+n)\,.
\end{equation}
We would like to know what is the global structure of this space. To this end we look for the group of discrete symmetries acting on the moduli that leave invariant the formulas \eqref{massform} and \eqref{levelmatch}, with suitable transformations on the quantum numbers $n,w,\pi$. In other words, we look for the T-duality group. 

A natural candidate for the T-duality group is the group of automorphisms of the momentum lattice. In the case of toroidal compactifications of the $E_8 \times E_8$ and the $Spin(32)/\mathbb{Z}_2$ heterotic strings, this correspondence is a standard result. It was in fact shown in \cite{Fraiman:2023cpa} that this correspondence also holds for toroidal compactifications of the rank 16 non-supersymmetric heterotic strings in ten dimensions, which is nothing but the case $n = 16$ in our family of heterotic strings. Let us briefly review this result in the context of generic $n$. 

First note that the full momentum lattice is the dual of $\Gamma^v_{1,n+1}$, as it contains $\Gamma^v_{1,n+1}$ itself as well as its three non-trivial conjugacy classes. However, the automorphism group of a lattice $L$ and its dual lattice $L^*$ are the same. Consider then the group of automorphisms of the lattice $\Gamma^v_{1,n+1} = \Gamma_{1,1}\oplus D_n$, denoted $O(\Gamma^v_{1,n+1})$. It is obvious that it mixes states in the spacetime vector class among themselves and by construction respects their quantization conditions. What we need to check is if it does the same for the other three spacetime classes. We may write any element of $\Gamma^o_{1,n+1}$ as an element of $\Gamma^v_{1,n+1}$ plus the vector $y$ with $n = w = 0$ and $\pi = (1,0^{n-1})$, and since $y^2 = 1$ it cannot be mapped to an element in any other class, hence $\Gamma^o_{1,n+1}$ is invariant as a class under $O(\Gamma^v_{1,n+1})$. Applying this analysis to the spinor classes it remains only a possibility that they are exchanged, but this is not an issue at all. Our theory is compactified on a circle and so it is not chiral, hence there is no physical distinction between the spinor and conjugate spinor classes. Finally, the group $O(\Gamma_{1,n+1}^v)$ is equivalent to $O(\Gamma_{1,n+1}^v\cup \Gamma_{1,n+1}^o)$, where $\Gamma_{1,n+1}^v \cup \Gamma_{1,n+1}^o \simeq \text{I}_{1,n+1}$ is the unique odd self-dual hyperbolic lattice,\footnote{This is easily seen by noting the isomorphism $D_n\cup(D_n+y) \simeq \mathbb{Z}^n$ where $y$ is the vector conjugacy class of $D_n$, and $\Gamma_{1,1}\oplus \mathbb{Z}^n \simeq \text{I}_{1,n+1}$.} hence $O(\Gamma_{1,n+1}^v)\simeq O(1,n+1;\mathbb{Z})$. 

\begin{figure}
	\centering
	\begin{tikzpicture}[scale = 1.25]
		\draw(1,-0.5)node{$\Sigma_{17}$};
		\draw(0,0)--(0.5,0.5);
		\draw(2,0)--(1.5,0.5);
		\draw(0,2)--(0.5,1.5);
		\draw(2,2)--(1.5,1.5);
		
		\draw[double, Red](0.5,0.5)--(1,0.75);
		\draw[double, Red](0.5,1.5)--(1,1.25);
		\draw[double, Red](1,0.75)--(1.5,1.5);
		\draw[double, Red, fill=white](1,1.25)--(1.5,0.5);
		\draw[double, Red](1,0.75)--(1,1.25);
		
		\draw(0,0)--(2,0)--(2,2)--(0,2)--(0,0);
		\draw[fill=white](0,0)circle(0.07);
		\draw[fill=white](0.5,0)circle(0.07);
		\draw[fill=white](1,0)circle(0.07);
		\draw[fill=white](1.5,0)circle(0.07);
		\draw[fill=white](2,0)circle(0.07);
		\draw[fill=white](0,0.5)circle(0.07);
		\draw[fill=white](0,1)circle(0.07);
		\draw[fill=white](0,1.5)circle(0.07);
		\draw[fill=white](2,0.5)circle(0.07);
		\draw[fill=white](2,1)circle(0.07);
		\draw[fill=white](2,1.5)circle(0.07);	
		\draw[fill=white](0,2)circle(0.07);
		\draw[fill=white](0.5,2)circle(0.07);
		\draw[fill=white](1,2)circle(0.07);
		\draw[fill=white](1.5,2)circle(0.07);
		\draw[fill=white](2,2)circle(0.07);
		
		\draw[fill=white](0.5,0.5)circle(0.07);
		\draw[fill=white](0.5,1.5)circle(0.07);
		\draw[fill=white](1.5,0.5)circle(0.07);
		\draw[fill=white](1.5,1.5)circle(0.07);
		\draw[Red, fill=white,double](1,1.25)circle(0.07);
		\draw[Red, fill=white,double](1,0.75)circle(0.07);
		\begin{scope}[shift={(3,0)}]
			\draw(1,-0.5)node{$\Sigma_{16}$};
			\draw[double, Red](0,0)--(0.75,0.75);
			\draw(2,0)--(1.5,0.5);
			\draw(0,2)--(0.5,1.5);
			\draw[double, Red](2,2)--(1.25,1.25);
			\draw[double, Red](0.75,0.75)--(1.25,1.25);
			
			\draw(0,0)--(2,0)--(2,2)--(0,2)--(0,0);
			\draw[fill=white](0,0)circle(0.07);
			\draw[fill=white](0.5,0)circle(0.07);
			\draw[fill=white](1,0)circle(0.07);
			\draw[fill=white](1.5,0)circle(0.07);
			\draw[fill=white](2,0)circle(0.07);
			\draw[fill=white](0,0.5)circle(0.07);
			\draw[fill=white](0,1)circle(0.07);
			\draw[fill=white](0,1.5)circle(0.07);
			\draw[fill=white](2,0.5)circle(0.07);
			\draw[fill=white](2,1)circle(0.07);
			\draw[fill=white](2,1.5)circle(0.07);	
			\draw[fill=white](0,2)circle(0.07);
			\draw[fill=white](0.5,2)circle(0.07);
			\draw[fill=white](1,2)circle(0.07);
			\draw[fill=white](1.5,2)circle(0.07);
			\draw[fill=white](2,2)circle(0.07);
			
			\draw[fill=white](0.5,1.5)circle(0.07);
			\draw[fill=white](1.5,0.5)circle(0.07);
			\draw[Red, fill=white,double](1.25,1.25)circle(0.07);
			\draw[Red, fill=white,double](0.75,0.75)circle(0.07);
		\end{scope}
		\begin{scope}[shift={(6,0)}]
			\draw(1,-0.5)node{$\Sigma_{15}$};
			\draw(2,0)--(1.5,0.5);
			\draw(0,2)--(0.5,1.5);
			
			\draw(0,0)--(2,0)--(2,2)--(0,2)--(0,0);
			
			\draw[Red, fill=white,double](1.5,2)--(2,2);
			\draw[Red, fill=white,double](2,1.5)--(2,2);
			
			\draw[fill=white](0,0)circle(0.07);
			\draw[fill=white](0.5,0)circle(0.07);
			\draw[fill=white](1,0)circle(0.07);
			\draw[fill=white](1.5,0)circle(0.07);
			\draw[fill=white](2,0)circle(0.07);
			\draw[fill=white](0,0.5)circle(0.07);
			\draw[fill=white](0,1)circle(0.07);
			\draw[fill=white](0,1.5)circle(0.07);
			\draw[fill=white](2,0.5)circle(0.07);
			\draw[fill=white](2,1)circle(0.07);
			\draw[fill=white](2,1.5)circle(0.07);	
			\draw[fill=white](0,2)circle(0.07);
			\draw[fill=white](0.5,2)circle(0.07);
			\draw[fill=white](1,2)circle(0.07);
			\draw[fill=white](1.5,2)circle(0.07);
			
			\draw[fill=white](0.5,1.5)circle(0.07);
			\draw[fill=white](1.5,0.5)circle(0.07);
			\draw[Red, fill=white,double](2,2)circle(0.07);
		\end{scope}
		\begin{scope}[shift={(9,0)}]
			\draw(1,-0.5)node{$\Sigma_{14}$};
			\draw(2,0)--(1.5,0.5);
			\draw(0,2)--(0.5,1.5);
			
			\draw(0,0)--(2,0)--(2,1.5);
			\draw(1.5,2)--(0,2)--(0,0);
			
			\draw[Red, fill=white,double](1.5,2)--(2,1.5);
			\draw[Red, fill=white,double](1,2)--(1.5,2);
			\draw[Red, fill=white,double](2,1)--(2,1.5);
			
			\draw[fill=white](0,0)circle(0.07);
			\draw[fill=white](0.5,0)circle(0.07);
			\draw[fill=white](1,0)circle(0.07);
			\draw[fill=white](1.5,0)circle(0.07);
			\draw[fill=white](2,0)circle(0.07);
			\draw[fill=white](0,0.5)circle(0.07);
			\draw[fill=white](0,1)circle(0.07);
			\draw[fill=white](0,1.5)circle(0.07);
			\draw[fill=white](2,0.5)circle(0.07);
			\draw[fill=white](2,1)circle(0.07);
			\draw[fill=white](2,1.5)circle(0.07);	
			\draw[fill=white](0,2)circle(0.07);
			\draw[fill=white](0.5,2)circle(0.07);
			\draw[fill=white](1,2)circle(0.07);
			\draw[fill=white](1.5,2)circle(0.07);
			
			\draw[fill=white](0.5,1.5)circle(0.07);
			\draw[fill=white](1.5,0.5)circle(0.07);
			\draw[Red, fill=white,double](1.5,2)circle(0.07);
			\draw[Red, fill=white,double](2,1.5)circle(0.07);
		\end{scope}
	\end{tikzpicture}
	\caption{Coxeter diagrams $\Sigma_{n+1}$ for the reflective T-duality group $O^r_{1,n+1}$ of $Spin(2n)$ non-supersymmetric heterotic strings on $S^1$ with $n = 16$ to $n = 13$. Regular nodes correspond to long roots in $\Gamma_{1,n+1}^v$ which furnish the enhancement $U(1)\to SU(2)$. Red nodes correspond to short roots belonging to the scalar class $\Gamma_{1,n+1}^o$ and furnish pairs of tachyons with lowest possible squared mass. }\label{fig:diag-leq17}
\end{figure}
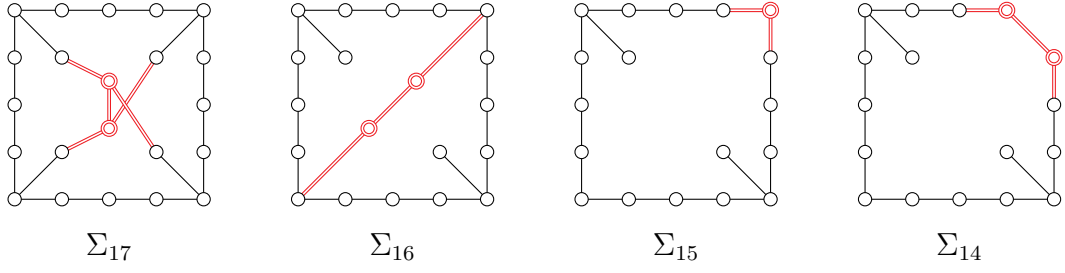

We conclude that for arbitrary $n$, the T-duality group is $O(\Gamma^v_{1,n+1}) \simeq O(1,n+1;\mathbb{Z})$, or $O_{1,n+1}$ for short. Fortunately for us, these groups were studied in detail by Vinberg \cite{Vinberg:1972} and Vinberg-Kaplinskaya \cite{Kaplinskaya:1978} a long time ago for $n \leq 18$; the cases $n = 19,...,22$ were worked out subsequently by Borcherds \cite{Borcherds:1987}. They take the form
\begin{equation}
	O_{1,n+1} = O^r_{1,n+1} \rtimes H_{n+1}\,,
\end{equation}
where $O^r_{1,n+1}$ is the maximal reflexive subgroup of $O_{1,n+1}$ and $H_{n+1}$ is the group of outer automorphisms of $\Gamma^v_{1,n+1}$.\footnote{The subscript $n+1$ denotes the dimension of the hyperbolic space we are working with, while $1,n+1$ is otherwise used to emphasize the signature of the corresponding object. Also note: the outer automorphism of $D_n$ is a reflection in $D_n^*$ generated by short roots, hence it is in $O^r_{1,n+1}$ and not in $H_{n+1}$.} The quotient
\begin{equation}
	\mathcal{P}_{n+1} = O^r_{1,n+1}\backslash\mathcal{M}_{1,n+1}\,
\end{equation}
defines a cover of the fundamental domain $O_{1,n+1}\backslash \mathcal{M}_{1,n+1}$ which is particularly convenient to work with. It is the fundamental cell of a group of reflections in hyperbolic space, known as a Coxeter polytope. These reflections  are finitely generated, such that the generators are in correspondence to a set of codimension 1 walls bounding $\mathcal{P}_{n+1}$. The way in which these walls intersect can be encoded in a Coxeter diagram $\Sigma_{n+1}$ as shown in Figure \ref{fig:diag-leq17} for $13\leq n \leq 16$. For supercritical strings these diagrams are more complicated, and we show the cases $n = 17,18$ in Figure \ref{fig:diag1819} (see \cite{Borcherds:1987} for higher $n$). The group of symmetries of the polytope $\mathcal{P}_{n+1}$ is equivalent to $H_{n+1}$ for $n \leq 18$.

At the locus of each wall represented by a white node, there occurs an $U(1) \to SU(2)$ symmetry enhancement in the left-moving gauge group. At the intersection of various walls, one reads off the corresponding symmetry enhancement from the Dynkin diagram they form. At walls represented by black nodes there is however a different kind of enhancement. Namely, a pair of tachyons acquire their lowest possible value of $m^2$. This corresponds to an ``enhancement" of a $Spin(2)$ symmetry with tachyons in the vector representation. The intersection of such a wall with the other ordinary walls is represented by a B-type Dynkin diagram, and must be read as an enhancement of a $Spin(2n)$ gauge symmetry together with $2n$ tachyons transforming in the vector representation. We should keep in mind that all of these enhancements lie in the NS sector and so the corresponding states have shifted $m^2$. The appearance of massless fermions is not directly encoded in these diagrams, but can be worked out separately. %The precise way in which these states appear will be dealt with in Section [].

\subsection{What we learn from the diagrams}
As already stated, the Coxeter diagram $\Sigma_{n+1}$ for a circle compactified $Spin(2n)$ heterotic string encodes the possible non-Abelian gauge symmetry enhancements in the moduli space. The exact values for the moduli $R$ and $A$ at which such enhancements occur can be obtained by specifying the root vectors corresponding to each node, i.e. by giving their values for $n,w$ and $\pi$. We leave the explanation for this procedure and the corresponding results to Section \ref{s:maximal}. Here we simply highlight some important features of our theories that these diagrams teach us. 

\subsubsection{Infinite distance limits}\label{ss:inf}
One of the main features of the diagrams $\Sigma_{n+1}$ is that they encode all the vertices of the polytope $\mathcal{P}_{n+1}$ as maximal Dynkin subdiagrams. These vertices represent maximal symmetry enhancements and can be at finite or infinite distance, corresponding respectively to ordinary or extended Dynkin diagrams. The infinite distance limits give decompactifications to the (non)critical heterotic strings of rank $n$ in the corresponding dimension $D$. For example, $\mathcal{P}_{17}$ has eight extended Dynkin subdiagrams reading
\begin{equation}\label{critical-inf}
	\begin{split}
		&~~~\widehat{E_8\oplus E_8}\,, ~~~~~ \widehat{D_{16}}\,, ~~~~~ \widehat{D_8 \oplus D_8}\,, ~~~~~ \widehat{A_{15}\oplus B_1}\,,\\
		&\widehat{E_7\oplus E_7 \oplus B_2}\,, ~~~~~ \widehat{D_{12}\oplus B_4}\,, ~~~~~ \widehat{E_8 \oplus B_8}\,, ~~~~~ \widehat{B_{16}}\,,
	\end{split}
\end{equation} 
corresponding to both the supersymmetric and the non-supersymmetric heterotic strings in ten dimensions with rank 16.\footnote{From the point of view of the lower dimensional theory, the gauge symmetry does correspond to an extended Dynkin diagram, i.e. it becomes affine, as shown for the supersymmetric case in \cite{Collazuol:2022jiy}. See also \cite{Collazuol:2022oey,Collazuol:2024kzl}.} The moduli space $\mathcal{M}_{1,17}$ thus interpolates between these eight theories, a fact that was already demonstrated in \cite{Ginsparg:1986wr}. Also note that by deleting the red nodes one easily restricts to infinite distance limits without tachyons. 

This interpolation between rank $n$ heterotic strings extends to the non-critical cases. For subcritical theories the subdiagrams are given by those in \eqref{critical-inf} by transforming $B_m \to B_{m-\ell}$ with $\ell = (10-D)/2$ when possible. For example, the diagram $\Sigma_{16}$ contains the five extended Dynkin subdiagrams
\begin{equation}\label{subdiag8}
	\widehat{A_{15}}\,, ~~~~~ \widehat{E_7 \oplus E_7 \oplus B_1}\,, ~~~~~ \widehat{D_{12}\oplus B_3}\,, ~~~~~ \widehat{E_8 \oplus B_7}\,, ~~~~~ \widehat{B_{15}}\,.
\end{equation}
For supercritical strings, the procedure is reversed. As one adds pairs of MW fermions $\lambda$, factors of type $B_m$ appear or are extended, and new tachyon-free theories appear (we mean tachyon-free up to $m^2$ shift). Indeed in $D = 12, 14$ there are tachyon-free heterotic strings with gauge algebras\footnote{In general, non-critical strings can be constructed by taking the left-moving internal part of the worldsheet CFT to be a chiral fermionic theory with $c = n$. This construction is nicely reviewed in \cite{BoyleSmith:2023xkd} for critical non-supersymmetric strings, and the classification of such chiral fermionic theories up to $c = 24$ is achieved in \cite{Hohn:2023auw} (see also \cite{Rayhaun:2023pgc}). For our $Spin(2n)$ theories and their T-duals, this classification is equivalent to that of odd self-dual lattices, which is much older (see \cite{King:2003} for $c \geq 25$).}
\begin{align}
	D = 12:& ~~~~~ \mathfrak{su}_{12}\oplus \mathfrak{e}_6\,,\\
	D = 14:& ~~~~~ \mathfrak{su}_{17}\oplus \mathfrak{su}_2\,, ~~~~~ \mathfrak{so}_{20}\oplus\mathfrak{e}_7 \oplus\mathfrak{su}_2\,, ~~~~~ 3\,\mathfrak{so}_{12}\,, ~~~~~ 2\mathfrak{su}_{10}\,,
\end{align}
and these can be seen as extended Dynkin subdiagrams of the corresponding Coxeter diagrams shown in Figure \ref{fig:diag1819}.

\subsubsection{How tachyon condensation affects the moduli space}
Tachyon condensation can dynamically reduce the number of target space dimensions of a heterotic string \cite{Hellerman:2006ff}, transforming for example a critical tachyonic theory to a subcritical tachyon-free one. This particular situation was studied in detail by Kaidi in \cite{Kaidi:2020jla}, following the work of Hellerman-Swanson \cite{Hellerman:2006ff,Hellerman:2007zz}. Here we consider the interplay between tachyon condensation and compactification on a circle as reflected on the Coxeter diagrams. We do not dwell on the physics of tachyon condensation but rather on the worldsheet theories it connects (up to a light-like linear dilaton in the initial setup.)

This formalism for tachyon condensation in heterotic strings requires the tachyon field $\mathcal{T}(X)$ to couple to left-moving free MW fermions $\lambda$ through a superpotential $W = \lambda :\mathcal{T}(X):$\,. This occurs when the tachyon has the lowest possible value of $m^2$, so that it corresponds to a $B_n$ subdiagram in the Coxeter diagram. If we wish only to condense a pair of tachyonic states then we can take them to correspond to one of the red nodes in the diagram, without loss of generality.\footnote{These tachyons correspond to norm 1 vectors in $\Gamma_{1,n+1}^o$, all of which lie in the same T-duality orbit. Two red nodes in $\Sigma_{n+1}$ are equivalent under outer automorphisms, i.e. symmetries of the diagram.} This process reduces the dimension by 2, and so it is naturally represented as a transformation of the Coxeter polytope $\mathcal{P}_{n+1}$ into $\mathcal{P}_n$. We can think of the corresponding Coxeter diagrams  as being connected through a series of transformations $\Sigma_{n+1} \to \Sigma_{n}$ representing the physical process of tachyon condensation. Indeed, this transformation proceeds by eliminating the tachyonic nodes and replacing ordinary nodes by new tachyonic ones in a way that is not entirely trivial.  

What is perhaps more important is the fact that, for $n \leq 16$, tachyon condensation connects one moduli space to another one in an unique way, and so this procedure commutes with circle compactification (for $n = 17,18$ there is a subtlety that we address below.) One may have anticipated this fact from the uniqueness of the moduli spaces themselves. In any case, this fact allows us to interpret any point in the moduli space of a subcritical string as resulting from the condensation of a tachyon in a higher dimensional theory, which is relevant for the study of heterotic branes.  

\subsubsection{Supercritical strings have two kinds of tachyons}
The groups $O_{1,18}$ and $O_{1,19}$ were studied by Vinberg and Kaplinskaya in \cite{Kaplinskaya:1978}. At a glance, a striking feature appears. Unlike the cases considered above, their associated diagrams $\Sigma_{18}$ and $\Sigma_{19}$ showcase two inequivalent kinds of walls corresponding to short roots. Let us work out what this means in the first case.
\begin{figure}
	\centering
	\begin{tikzpicture}[scale = 1.25]
		\draw(2,-0.5)node{$\Sigma_{18}$};
			\draw(0,0)--(4,0)--(2,4)--(0,0);
			\draw(2,1.5)--(2,4);
			\draw(2,1.5)--(0,0);
			\draw(2,1.5)--(4,0);
			%Outer triangle			
			\draw[fill=white](0,0)circle(0.07);
			\draw[fill=white](1,0)circle(0.07);
			\draw[fill=white](2,0)circle(0.07);
			\draw[fill=white](3,0)circle(0.07);
			\draw[fill=white](4,0)circle(0.07);
			%\draw[fill=white](0.5,1)circle(0.07);
			%\draw[fill=white](1,2)circle(0.07);
			\draw[fill=white](1.5,3)circle(0.07);
			\draw[fill=white](3.5,1)circle(0.07);
			%\draw[fill=white](3,2)circle(0.07);
			\draw[fill=white](2.5,3)circle(0.07);	
			\draw[fill=white](2,4)circle(0.07);
			%Inner affine E6
			\draw[fill=white](2,1.5)circle(0.07);
			\draw[fill=white](2,2.125)circle(0.07);
			\draw[fill=white](2,2.75)circle(0.07);
			\draw[fill=white](2,3.375)circle(0.07);
			\draw[fill=white](1.5,1.125)circle(0.07);
			\draw[fill=white](1,0.75)circle(0.07);
			\draw[fill=white](0.5,0.375)circle(0.07);
			%\draw[fill=white](2.5,1.125)circle(0.07);
			%\draw[fill=white](3,0.75)circle(0.07);
			\draw[fill=white](3.5,0.375)circle(0.07);
			%Tachyons
			\draw[Red, fill=white,double](2,1.125)--(1,2);
			\draw[Red, fill=white,double](2,1.125)--(3,0.75);
			\draw[Blue, fill=white,thick,dotted](2.65,0.5)--(3,2);
			\draw[Blue, fill=white,thick,dotted](2.65,0.5)--(2.5,1.125);
			\draw[Blue, fill=white,thick,dotted](2.65,0.5)--(0.5,1);
			\draw[fill=white](1,2)circle(0.07);
			\draw[fill=white](3,0.75)circle(0.07);
			\draw[fill=white](3,2)circle(0.07);
			\draw[fill=white](2.5,1.125)circle(0.07);
			\draw[fill=white](0.5,1)circle(0.07);
			\draw[Red, fill=white,double](2,1.125)circle(0.07);
			\draw[Blue, fill=white,double](2.65,0.5)circle(0.07);
			\begin{scope}[shift={(7,0)}]
			\draw(1.3,-0.5)node{$\Sigma_{19}$};
			%Black lines
			\draw(2.5,0)--(0,0)--(-0.77,2.37)--(1.25,3.84)--(3.27,2.37)--(2.5,0);
			\draw(0.47,0.64)--(1.25,3.04)--(2.03,0.64)--(0,2.12)--(2.5,2.12)--(0.47,0.64);
			\draw(0,0)--(0.47,0.64);
			\draw(2.5,0)--(2.03,0.64);
			\draw(-0.77,2.37)--(0,2.12);
			\draw(3.27,2.37)--(2.5,2.12);
			\draw(1.25,3.84)--(1.25,3.04);
			%Red and blue lines
			\draw[Red, fill=white,double](2.5,1.75)--(1.25,3.44);
			\draw[Red, fill=white,double](2.5,1.75)--(1.25,2.12);
			\draw[Red, fill=white,double](2.5,1.75)--(1.25,0);
			\draw[Blue, fill=white,thick,dotted](0.2,1.35)--(-0.38,1.18);
			\draw[Blue, fill=white,thick,dotted](0.2,1.35)--(1.46,1.78);
			\draw[Blue,thick,dotted](1.46,1.78) arc[start angle=-144, end angle=0, radius=0.2];
			\draw[Blue,thick,dotted](1.85,1.9)--(2.5,2.12);
			\draw[Blue,thick,dotted](0.2,1.35)--(1.25,3.44);
			\draw[Blue, fill=white,thick,dotted](0.2,1.35)--(2.27,0.32);
			%Outer nodes
			\draw[fill=white](0,0)circle(0.07);
			\draw[fill=white](1.25,0)circle(0.07);
			\draw[fill=white](2.5,0)circle(0.07);
			\draw[fill=white](-0.38,1.18)circle(0.07);
			\draw[fill=white](-0.77,2.37)circle(0.07);
			\draw[fill=white](2.88,1.18)circle(0.07);
			\draw[fill=white](3.27,2.37)circle(0.07);
			\draw[fill=white](0.24,3.11)circle(0.07);
			\draw[fill=white](1.25,3.84)circle(0.07);
			\draw[fill=white](2.26,3.11)circle(0.07);
			%Third layer from outside
			\draw[fill=white](0.47,0.64)circle(0.07);
			\draw[fill=white](2.03,0.64)circle(0.07);
			\draw[fill=white](0,2.12)circle(0.07);
			\draw[fill=white](2.5,2.12)circle(0.07);
			\draw[fill=white](1.25,3.04)circle(0.07);
			%Second layer from outside
			\draw[fill=white](0.23,0.32)circle(0.07);
			\draw[fill=white](2.27,0.32)circle(0.07);
			\draw[fill=white](-0.38,2.24)circle(0.07);
			\draw[fill=white](2.88,2.24)circle(0.07);
			\draw[fill=white](1.25,3.44)circle(0.07);
			%Inned pentagon nodes
			\draw[fill=white](1.25,2.12)circle(0.07);
			\draw[fill=white](0.86,1.84)circle(0.07);
			\draw[fill=white](1.64,1.84)circle(0.07);
			\draw[fill=white](1,1.39)circle(0.07);
			\draw[fill=white](1.5,1.39)circle(0.07);
			%Red and blue nodes
			\draw[Red, fill=white,double](2.5,1.75)circle(0.07);
			\draw[Blue, fill=white,double](0.2,1.35)circle(0.07);
			\end{scope}
	\end{tikzpicture}
	\caption{Coxeter diagrams for the $D = 12$ and $D = 14$ supercritical heterotic string on $S^1$. Not all red and blue nodes, nor their links, are shown, although they are unique up to diagram automorphisms. For $D = 12$, condensation of tachyons associated to (red) blue nodes yields a (non)supersymmetric theory. More details regarding these diagrams can be found in \cite{Guglielmetti:2015}.}\label{fig:diag1819}
\end{figure}
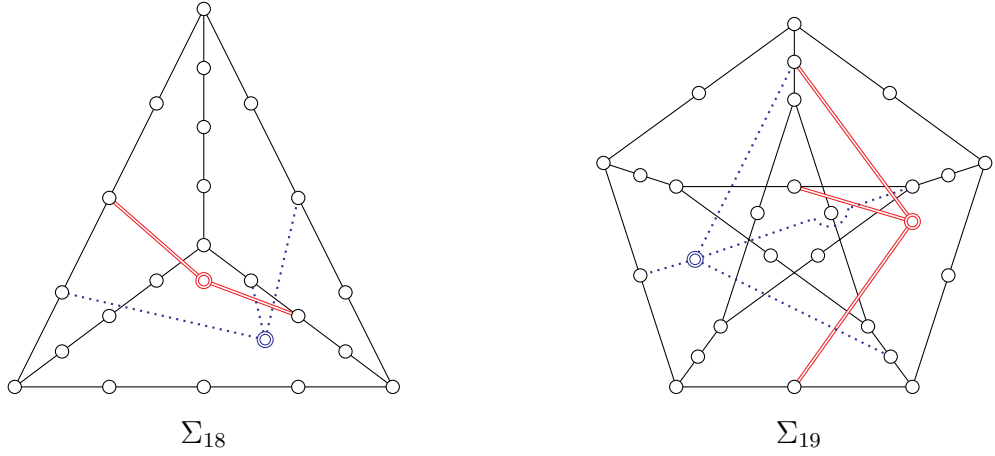

The Coxeter diagram $\Sigma_{18}$ is shown in Figure \ref{fig:diag1819}. The long roots form a tetrahedron which naturally results from replacing the short roots in $\Sigma_{17}$ by two disconnected long roots. There are three short roots \textit{of the first kind}, which are analogous to those found in $\Sigma_{p+1 \leq 17}$. They are linked to the long roots with double lines, and the intersections of their corresponding walls occur at infinite distance. We thus expect that tachyon condensation as explained above leads to the theory corresponding to $\Sigma_{17}$. On the other hand, we find 12 new short roots \textit{of the second kind}, which are linked in a different manner. A physically sensible expectation is that such roots correspond to tachyons which condense to a critical \textit{supersymmetric} theory. 

Here we draw only one representative for each inequivalent short root. As for the previous cases $\Sigma_{n+1 \leq 17}$, the red nodes form an orbit under the symmetry group of the full diagram, and can be easily seen to add up to three using a $\mathbb{Z}_3$ subgroup of $H_{18}\simeq S_4$. Similarly, the blue nodes form a 12-element orbit under $H_{18}$. Tachyon condensation selects out the orthogonal complement of a $D_1$ sublattice in $\Gamma_{1,18}^v$, and we expect there to be two inequivalent choices. These can indeed be read off from the two presentations of $\Gamma_{1,18}^v$, 
\begin{equation}
	\Gamma_{1,1}\oplus [D_{16}\oplus \textcolor{Red}{D_1}]^{(v,v)} \simeq \Gamma_{1,1}\oplus E_8 \oplus E_8 \oplus \textcolor{Blue}{D_1}\,,
\end{equation}
where $(v,v)$ is a gluing vector such that $[D_{16}\oplus D_1]^{(v,v)} = D_{17}$. In the first case, the orthogonal complement of $D_1$ gives $\Gamma_{1,17}^v$, while in the second it gives the Narain lattice of supersymmetric heterotic strings on $S^1$. Incidentally, if one removes all the tachyonic nodes in $\Sigma_{18}$ as well as the three nodes connected to the blue node, there remains the Coxeter diagram of $\Gamma_{1,1}\oplus E_8\oplus E_8$. The reason is that a given blue node is connected to every other blue node as well as every red node; removing all the connected nodes leaves out a set of generating roots for the supersymmetric Narain lattice, which is the diagramatic equivalent of taking the orthogonal complement of the corresponding $D_1$.

The diagram $\Sigma_{19}$ also contains short roots of the first and second kind. Condensing a pair of tachyons of any type will inevitably lead to $\Sigma_{18}$ as there is no $D = 12$ supersymmetric heterotic string. The (primitive) lattice embedding $D_1 \hookrightarrow \Gamma_{1,19}^v$ must be unique. Presumably, the difference between the types of nodes in $\Sigma_{19}$ accounts for the inequivalent ways of condensing \textit{four tachyons}, which can indeed lead to the supersymmetric theory in ten dimensions. The structure of this diagram is more intricate than $\Sigma_{18}$, containing in total five short roots of the first kind and twenty of the second kind. One can carry out this construction in a controlled manner up to $\Sigma_{23}$, see \cite{Borcherds:1987}.

\subsection{Note on 2D strings}\label{ss:2dstrings}
The case of $D = 2$ subcritical theories is special in that the space-like direction cannot be compactified due to the linear dilaton profile. One may however compactify the time direction, obtaining a moduli space as before. This setup was studied some time ago in \cite{Davis:2005qe,Davis:2005qi,Seiberg:2005nk}, but the T-duality group remained undetermined. It was sensibly conjectured to be $O(1,13;\mathbb{Z})$ and our results imply that this is indeed the case.\footnote{We do note that in the physics literature the notation $O(p,q)$ is usually abused in that there is no specification of the metric with respect to which the elements of the group are orthogonal. In our case we do mean orthogonal matrices with respect to the Minkowskian metric, which act as automorphisms of the odd self-dual lattices. Such lattices appear naturally in the covariant formulation of \cite{Lerche:1986he} used in \cite{Davis:2005qe}.} The mass formula and level matching conditions for the spectrum follow the same structure as the spacelike compactifications in higher dimensions due to Lorentz invariance, so that the type of T-duality group generalizes to this case even if we have not written down the partition function explicitly.  

The group $O(1,13;\mathbb{Z})$ is purely reflective, and so the moduli space is exactly the Coxeter polytope $\mathcal{P}_{13}$. In going from $\Sigma_{14}$ to $\Sigma_{13}$ the $\mathbb{Z}_2$ outer automorphism is broken. The Coxeter diagram $\Sigma_{13}$ takes the following form \cite{Vinberg:1972}:\footnote{The labeling of nodes follows the general form employed in Section \ref{s:maximal}.}
\begin{eqnarray}
		\begin{tikzpicture}[scale = 1.25]
			\draw(0.5,0)--(0.5,1);
			\draw(4.5,0)--(4.5,0.5);
			
			\draw(0,0)--(5,0);
			
			\draw[Red, fill=white,double](4.5,0.5)--(4.5,1);
			
			\draw[fill=white](0,0)circle(0.07)node[below]{\scriptsize{\textcolor{darkgray}{1}}};
			\draw[fill=white](0.5,0)circle(0.07)node[below]{\scriptsize{\textcolor{darkgray}{2}}};
			\draw[fill=white](1,0)circle(0.07)node[below]{\scriptsize{\textcolor{darkgray}{3}}};
			\draw[fill=white](1.5,0)circle(0.07)node[below]{\scriptsize{\textcolor{darkgray}{4}}};
			\draw[fill=white](2,0)circle(0.07)node[below]{\scriptsize{\textcolor{darkgray}{5}}};
			\draw[fill=white](2.5,0)circle(0.07)node[below]{\scriptsize{\textcolor{darkgray}{6}}};
			\draw[fill=white](3,0)circle(0.07)node[below]{\scriptsize{\textcolor{darkgray}{7}}};
			\draw[fill=white](3.5,0)circle(0.07)node[below]{\scriptsize{\textcolor{darkgray}{8}}};
			\draw[fill=white](4,0)circle(0.07)node[below]{\scriptsize{\textcolor{darkgray}{9}}};
			\draw[fill=white](4.5,0)circle(0.07)node[below]{\scriptsize{\textcolor{darkgray}{10}}};
			\draw[fill=white](5,0)circle(0.07)node[below]{\scriptsize{\textcolor{darkgray}{$\gamma$}}};
			
			\draw[fill=white](0.5,0.5)circle(0.07)node[left]{\scriptsize{\textcolor{darkgray}{0}}};
			\draw[fill=white](4.5,0.5)circle(0.07)node[right]{\scriptsize{\textcolor{darkgray}{11}}};
			\draw[fill=white](0.5,1)circle(0.07)node[left]{\scriptsize{\textcolor{darkgray}{$\beta$}}};
			\draw[Red, fill=white,double](4.5,1)circle(0.07)node[right]{\scriptsize{\textcolor{darkgray}{12}}};
		\end{tikzpicture}
\end{eqnarray} 
By deleting the pairs $(\beta,12)$, $(\beta,\gamma)$ we obtain the diagrams $\widehat{D_{16}}$ and $\widehat{B_{16}}$ corresponding to the two heterotic theories referred to as HO and THO in \cite{Seiberg:2005nk}. The first has gauge group $G = Spin(24)/\mathbb{Z}_2$, while the second has $G = Spin(24)$, with 24 tachyonic states in the vector representation of $G$ becoming massless due to the linear dilaton shift. Deleting node $8$ results in the diagram $\widehat{E_8\oplus B_4}$ corresponding to the heterotic $E_8\times Spin(8)$ theory likewise referred to as HE. The interpolation between the three heterotic theories is hence manifest.

\section{Maximal symmetry enhancements}\label{s:maximal}
In this section we explain how the moduli for all maximal symmetry enhancements can be extracted from the Coxeter diagrams for each non-critical circle compactification. We then show how the different enhancements are related across dimensions through tachyon condensation, allowing to extract the spectral data for a $D$-dimensional theory from that of the $(D+2)$-dimensional one. In particular, the data for the critical case has been obtained in \cite{Fraiman:2023cpa}, hence the data for all subcritical theories can be readily obtained. 

\subsection{Extracting the data from the diagrams}
To work with a Coxeter diagram we must first choose a presentation for the roots corresponding to each node. We will use the $Spin(2n)$ heterotic frame, which easily generalizes across dimensions. 

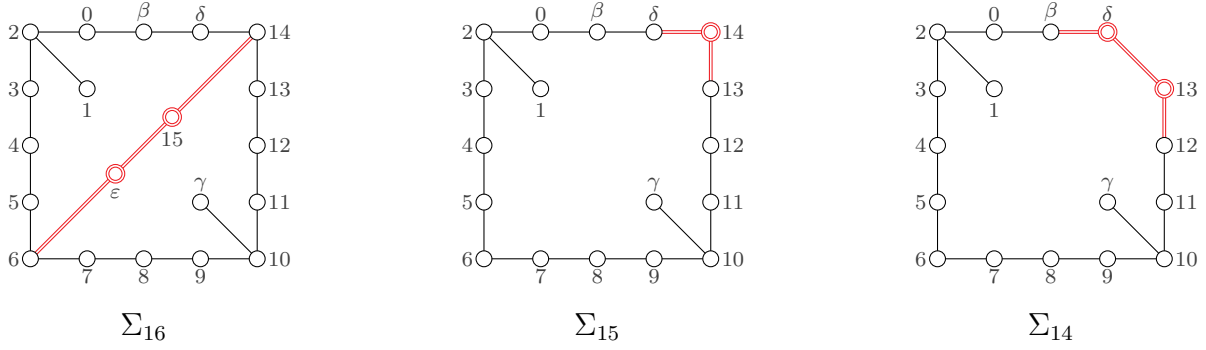
\begin{figure}
	\centering
	\begin{tikzpicture}[scale = 1.5]
		\draw(1,-0.6)node{$\Sigma_{16}$};
		\draw[double, Red](0,0)--(0.75,0.75);
		\draw(2,0)--(1.5,0.5);
		\draw(0,2)--(0.5,1.5);
		\draw[double, Red](2,2)--(1.25,1.25);
		\draw[double, Red](0.75,0.75)--(1.25,1.25);
		
		\draw(0,0)--(2,0)--(2,2)--(0,2)--(0,0);
		\draw[fill=white](0,0)circle(0.07)node[left]{\scriptsize{\textcolor{darkgray}{6}}};
		\draw[fill=white](0.5,0)circle(0.07)node[below]{\scriptsize{\textcolor{darkgray}{7}}};
		\draw[fill=white](1,0)circle(0.07)node[below]{\scriptsize{\textcolor{darkgray}{8}}};
		\draw[fill=white](1.5,0)circle(0.07)node[below]{\scriptsize{\textcolor{darkgray}{9}}};
		\draw[fill=white](2,0)circle(0.07)node[right]{\scriptsize{\textcolor{darkgray}{10}}};
		\draw[fill=white](0,0.5)circle(0.07)node[left]{\scriptsize{\textcolor{darkgray}{5}}};
		\draw[fill=white](0,1)circle(0.07)node[left]{\scriptsize{\textcolor{darkgray}{4}}};
		\draw[fill=white](0,1.5)circle(0.07)node[left]{\scriptsize{\textcolor{darkgray}{3}}};
		\draw[fill=white](2,0.5)circle(0.07)node[right]{\scriptsize{\textcolor{darkgray}{11}}};
		\draw[fill=white](2,1)circle(0.07)node[right]{\scriptsize{\textcolor{darkgray}{12}}};
		\draw[fill=white](2,1.5)circle(0.07)node[right]{\scriptsize{\textcolor{darkgray}{13}}};	
		\draw[fill=white](0,2)circle(0.07)node[left]{\scriptsize{\textcolor{darkgray}{2}}};
		\draw[fill=white](0.5,2)circle(0.07)node[above]{\scriptsize{\textcolor{darkgray}{0}}};
		\draw[fill=white](1,2)circle(0.07)node[above]{\scriptsize{\textcolor{darkgray}{$\beta$}}};
		\draw[fill=white](1.5,2)circle(0.07)node[above]{\scriptsize{\textcolor{darkgray}{$\delta$}}};
		\draw[fill=white](2,2)circle(0.07)node[right]{\scriptsize{\textcolor{darkgray}{14}}};
		
		\draw[fill=white](0.5,1.5)circle(0.07)node[below=0.02in]{\scriptsize{\textcolor{darkgray}{1}}};
		\draw[fill=white](1.5,0.5)circle(0.07)node[above]{\scriptsize{\textcolor{darkgray}{$\gamma$}}};
		\draw[Red, fill=white,double](1.25,1.25)circle(0.07)node[below=0.02in]{\scriptsize{\textcolor{darkgray}{15}}};
		\draw[Red, fill=white,double](0.75,0.75)circle(0.07)node[below=0.02in]{\scriptsize{\textcolor{darkgray}{$\varepsilon$}}};
		\begin{scope}[shift={(4,0)}]
			\draw(1,-0.6)node{$\Sigma_{15}$};
			\draw(2,0)--(1.5,0.5);
			\draw(0,2)--(0.5,1.5);
			
			\draw(0,0)--(2,0)--(2,2)--(0,2)--(0,0);
			
			\draw[Red, fill=white,double](1.5,2)--(2,2);
			\draw[Red, fill=white,double](2,1.5)--(2,2);
			
			\draw[fill=white](0,0)circle(0.07)node[left]{\scriptsize{\textcolor{darkgray}{6}}};
			\draw[fill=white](0.5,0)circle(0.07)node[below]{\scriptsize{\textcolor{darkgray}{7}}};
			\draw[fill=white](1,0)circle(0.07)node[below]{\scriptsize{\textcolor{darkgray}{8}}};
			\draw[fill=white](1.5,0)circle(0.07)node[below]{\scriptsize{\textcolor{darkgray}{9}}};
			\draw[fill=white](2,0)circle(0.07)node[right]{\scriptsize{\textcolor{darkgray}{10}}};
			\draw[fill=white](0,0.5)circle(0.07)node[left]{\scriptsize{\textcolor{darkgray}{5}}};
			\draw[fill=white](0,1)circle(0.07)node[left]{\scriptsize{\textcolor{darkgray}{4}}};
			\draw[fill=white](0,1.5)circle(0.07)node[left]{\scriptsize{\textcolor{darkgray}{3}}};
			\draw[fill=white](2,0.5)circle(0.07)node[right]{\scriptsize{\textcolor{darkgray}{11}}};
			\draw[fill=white](2,1)circle(0.07)node[right]{\scriptsize{\textcolor{darkgray}{12}}};
			\draw[fill=white](2,1.5)circle(0.07)node[right]{\scriptsize{\textcolor{darkgray}{13}}};
			\draw[fill=white](0,2)circle(0.07)node[left]{\scriptsize{\textcolor{darkgray}{2}}};
			\draw[fill=white](0.5,2)circle(0.07)node[above]{\scriptsize{\textcolor{darkgray}{0}}};
			\draw[fill=white](1,2)circle(0.07)node[above]{\scriptsize{\textcolor{darkgray}{$\beta$}}};
			\draw[fill=white](1.5,2)circle(0.07)node[above]{\scriptsize{\textcolor{darkgray}{$\delta$}}};
			
			\draw[fill=white](0.5,1.5)circle(0.07)node[below=0.02in]{\scriptsize{\textcolor{darkgray}{1}}};
			\draw[fill=white](1.5,0.5)circle(0.07)node[above]{\scriptsize{\textcolor{darkgray}{$\gamma$}}};
			\draw[Red, fill=white,double](2,2)circle(0.07)node[right]{\scriptsize{\textcolor{darkgray}{14}}};
		\end{scope}
		\begin{scope}[shift={(8,0)}]
			\draw(1,-0.6)node{$\Sigma_{14}$};
			\draw(2,0)--(1.5,0.5);
			\draw(0,2)--(0.5,1.5);
			
			\draw(0,0)--(2,0)--(2,1.5);
			\draw(1.5,2)--(0,2)--(0,0);
			
			\draw[Red, fill=white,double](1.5,2)--(2,1.5);
			\draw[Red, fill=white,double](1,2)--(1.5,2);
			\draw[Red, fill=white,double](2,1)--(2,1.5);
			
			\draw[fill=white](0,0)circle(0.07)node[left]{\scriptsize{\textcolor{darkgray}{6}}};
			\draw[fill=white](0.5,0)circle(0.07)node[below]{\scriptsize{\textcolor{darkgray}{7}}};
			\draw[fill=white](1,0)circle(0.07)node[below]{\scriptsize{\textcolor{darkgray}{8}}};
			\draw[fill=white](1.5,0)circle(0.07)node[below]{\scriptsize{\textcolor{darkgray}{9}}};
			\draw[fill=white](2,0)circle(0.07)node[right]{\scriptsize{\textcolor{darkgray}{10}}};
			\draw[fill=white](0,0.5)circle(0.07)node[left]{\scriptsize{\textcolor{darkgray}{5}}};
			\draw[fill=white](0,1)circle(0.07)node[left]{\scriptsize{\textcolor{darkgray}{4}}};
			\draw[fill=white](0,1.5)circle(0.07)node[left]{\scriptsize{\textcolor{darkgray}{3}}};
			\draw[fill=white](2,0.5)circle(0.07)node[right]{\scriptsize{\textcolor{darkgray}{11}}};
			\draw[fill=white](2,1)circle(0.07)node[right]{\scriptsize{\textcolor{darkgray}{12}}};
			\draw[fill=white](0,2)circle(0.07)node[left]{\scriptsize{\textcolor{darkgray}{2}}};
			\draw[fill=white](0.5,2)circle(0.07)node[above]{\scriptsize{\textcolor{darkgray}{0}}};
			\draw[fill=white](1,2)circle(0.07)node[above]{\scriptsize{\textcolor{darkgray}{$\beta$}}};
			
			\draw[fill=white](0.5,1.5)circle(0.07)node[below=0.02in]{\scriptsize{\textcolor{darkgray}{1}}};
			\draw[fill=white](1.5,0.5)circle(0.07)node[above]{\scriptsize{\textcolor{darkgray}{$\gamma$}}};
			\draw[Red, fill=white,double](1.5,2)circle(0.07)node[above]{\scriptsize{\textcolor{darkgray}{$\delta$}}};
			\draw[Red, fill=white,double](2,1.5)circle(0.07)node[right]{\scriptsize{\textcolor{darkgray}{13}}};
		\end{scope}
	\end{tikzpicture}
	\caption{Labeled Coxeter diagrams for the moduli spaces of subcritical strings compactified on $S^1$, with total dimension $D = 8,6,4$. See Section \ref{ss:2dstrings} for the case $D = 2$.}\label{fig:diag-leq17-numbered}
\end{figure}

There is a $B_n$ chain corresponding to the gauge bundle, whose simple roots may be written as
\begin{equation}
	\alpha_1 = (1,-1,0^{n-2})\,, ~~~~~...\,,~~~~~ \alpha_{n-1} = (0^{n-2},1,-1)\,, ~~~~~ \alpha_n = (0^{n-1},1)\,.
\end{equation}
We embed them into the charge sublattice $\Gamma_{1,n+1}^v \cup \Gamma_{1,n+1}^o \simeq \text{I}_{1,n+1}$ as vectors 
\begin{equation}
	\varphi_i = \ket{0,0;\alpha_i}\,, ~~~~~i = 1,...,n\,,
\end{equation}
and extend the diagram to $\widehat B_n$ by adding the lowest root with non-trivial KK momentum
\begin{equation}
	\varphi_0 = \ket{-1,0;-1,-1,0^{n-2}}\,.
\end{equation}
For generic $n$ there are two more extensions given by the vectors
\begin{equation}
	\varphi_\beta = \ket{1,1;0^n}\,, ~~~~~ \varphi_\gamma = \ket{-2,2;-1^{10},0^{n-10}}\,.
\end{equation}
For $n = 12$ the set $\{\varphi_0,...,\varphi_n,\varphi_\beta,\varphi_\gamma\}$ furnishes the whole Coxeter diagram. For higher $n$ we add the vectors
\begin{equation}
	\begin{split}
		\underline{n = 13:}~~~~~~~~~~& \varphi_\delta = \ket{-3,2;-1^{13}}\,,\\
		\underline{n = 14:}~~~~~~~~~~& \varphi_\delta = \ket{-3,2;-1^{14}}\,,\\
		\underline{n = 15:}~~~~~~~~~~& \varphi_\delta = \ket{-3,2;-1^{14},0}\,, ~~~~~ \varphi_\varepsilon = \ket{-4,4;-2^6,-1^9}\,.
	\end{split}
\end{equation}
The way in which these vectors give rise to the Coxeter diagrams is given in Figure \ref{fig:diag-leq17-numbered}.

From the mass formula \eqref{massform}, the locations of the walls are found by setting $p_R = 0$ in \eqref{momentum-pR} after plugging in the quantum numbers for the corresponding root:
\begin{equation}
	\begin{split}
		\varphi_i:& ~~~~~~~~~~ A_i - A_{i+1} = 0\,, ~~~~~ i = 1,...,n-1\,,\\
		\varphi_n:& ~~~~~~~~~~ A_n = 0\,,\\
		\varphi_\beta:& ~~~~~~~~~~ R^2 + \tfrac12 A^2 = 1\,,\\
		\varphi_\gamma:& ~~~~~~~~~~ 2R^2 + A^2 = \sum_{i = 1}^{10}A_i - 2\,,\\
		\varphi_\delta:& ~~~~~~~~~~ 2R^2 + A^2 = \sum_{i =1}^{m}A_i-3\,, ~~~~~ m = \begin{cases}13 & n = 13\\ 14 & n = 14,15	\end{cases}\,,\\
		\varphi_\epsilon:& ~~~~~~~~~~ 4R^2 + 2A^2 = 2\sum_{i = 1}^6 A_i + \sum_{i = 7}^{15}A_i - 4\,.
	\end{split}
\end{equation}
Symmetry enhancements are then obtained by selecting a Dynkin subdiagram of the Coxeter diagram and then fixing the constraints associated to the corresponding nodes. Similarly, decompactification limits are obtained by fixing the constraints associated to an affine Dynkin subdiagram. 

\subsection{Transforming the spectra through tachyon condensation}
As we have explained, every point in the moduli space of some heterotic string can be interpreted as resulting from tachyon condensation of another in higher dimensions. In the case of circle compactifications these moduli spaces are unique, and this implies for example that every maximal symmetry enhancement contained in the diagram $\Sigma_{16}$ can be obtained from one in $\Sigma_{17}$ for which there is a pair of tachyons. 

The question is then what is the effect of tachyon condensation at the level of the light spectrum of the higher dimensional theory. From the results of \cite{Kaidi:2020jla} we expect that the light bosonic states acquire a mass shift while massless fermionic states remain massless. Let us then explain how this expectation is realized in our language:
\begin{itemize}
	\item \underline{Bosons:} Bosonic states have electric charge lying in the lattice $\Gamma^{v}_{1,n+1}\cup \Gamma^o_{1,n+1} \simeq \text{I}_{1,n+1}$. A generic region in moduli space with a pair of maximally tachyonic states is given by the polarization $\text{I}_{1,n+1} = \text{I}_{1,n}\oplus \mathbb{Z}$, with the tachyonic states having charge $\pm 1$ in $\mathbb{Z}$. The bosonic states in the lower dimensional theory have charges in $\text{I}_n$ and will simply correspond to those in the original theory which are neutral under the $Spin(2)$ associated to the tachyons, with their mass shifted as in eq. \eqref{massform}.
	
	\item \underline{Fermions:} Writing the vector class as $\Gamma^v_{1,n+1} = \Gamma_{1,1}\oplus D_n$, the spinor classes all have $\pi \in D_n+s$ or $\pi \in D_n+c$, hence for any enhancement of the gauge symmetry group to $Spin(2n)$ with tachyons in the vector representation, every fermionic state in the spectrum is charged under $Spin(2n)$ in the spinor or co-spinor representation. In the case $n = 1$ corresponding to the situation being considered now, the squared mass of every fermionic state receives a contribution $P_{L,t}^2 \in (2\mathbb{Z}+1)^2/4$ from its $Spin(2)$ charge; for massless states this is $P_{L,t}^2 = 1/4$. This contribution is absent in the lower dimensional theory, and since $\delta_D - \delta_{D-2} = -1/4$, the mass of the corresponding fermionic states is invariant when it is zero. More precisely, there are towers of fermionic states with various $Spin(2)$ charge which ``collapse" to massless states in the lower dimensional theory; a similar situation holds for massive states.
\end{itemize}

What we learn from this analysis is that the spectrum of every maximal enhancement in a $D$-dimensional theory can be obtained in a simple way from a maximal enhancement in a $(D+2)$-dimensional theory with a pair of extremal tachyons. In particular we know the complete list of maximal enhancements in the case $D = 10$ \cite{Fraiman:2023cpa}, and from it we can obtain every enhancement in the subcritical theories. The data obtained in this way will be exactly equivalent to that obtained from the Coxeter diagrams as described above. The latter method offers the advantage of furnishing the values of the moduli for which the enhancements occur, which can be put into the partition function \eqref{ZDS1} to obtain the degeneracies for all massive levels. In the next section we will look at some enhancements which have a special interpretation in terms of heterotic branes.

\section{Applications to heterotic branes}\label{s:branes}
In this section use the interpretation of non-critical strings as furnishing linear dilaton backgrounds of branes applied to the heterotic string (see e.g. \cite{Murthy:2006eg}). We will first consider the subcritical strings arising from tachyon condensation of the critical theory on a circle at special points corresponding to Scherk-Schwarz reductions at self-dual radius, which turn out to correspond to CHS models describing pairs of coincident NS5 branes. We then show how in the case of the $Spin(32)/\mathbb{Z}_2$ heterotic string this background is dual to that of the non-supersymmetric 6-brane of \cite{Kaidi:2023tqo} compactified on a circle with suitable Wilson lines, a result which slightly refines some of those obtained in \cite{Israel:2004vv} at the level of sphere sigma models. A similar relationship is seen to hold between the non-supersymmetric 4-brane on $T^2$ and an intersection of two pairs of NS5-branes in the $E_8 \times E_8$ string, but only seems valid at low energies.

\subsection{A pair of NS5's from tachyon condensation}\label{ss:ns5s}

Recall that a Scherk-Schwarz reduction is realized by compactifying a theory, say the $E_8\times E_8$ heterotic string, on a circle $S^1$ with a $(-1)^F$ holonomy, where $F$ is the spacetime fermion number. It is well known that as the radius $R$ of the circle becomes smaller, two winding states
\begin{equation}
	\pm \ket{n = 1, w = \tfrac12; \pi = 0}
\end{equation}
become massless at a critical radius $R = R_H$ below which they are tachyonic, signaling the so-called Hagedorn phase transition \cite{Atick:1988si} in the analogous time compactification. For heterotic strings, T-duality implies that the tachyons actually acquire their lowest possible squared mass at the self-dual radius $R_\text{sd}$, after which there is a presumed second Hagedorn radius $R_H'$. What is important for us is the fact that at $R = R_\text{sd}$ the pair of winding tachyons have lowest possible $m^2$ and so their condensation can be studied using the machinery of \cite{Hellerman:2006ff}.\footnote{It is also possible to deform the theory through condensation of tachyons with lower $|m^2|$, in which case the deformations are relevant, see \cite{Saxena:2024eil}. It would be interesting to transpose that analysis to the present setup.}

A second important point is that at $R = R_\text{sd}$ two extra winding states
\begin{equation}
	\pm \ket{n = 1, w = -\tfrac12;0}
\end{equation}
become massless, enhancing the graviphoton $U(1)$ to an $SU(2)$ at level 2. It follows that the contribution from the compact direction to the worldsheet CFT consists of a pair of free left-moving MW fermions $\lambda,\lambda'$ associated to the tachyons and a triplet of free right-moving MW fermions furnishing the $SU(2)_2$ current. 

Turning on a suitable light-like linear dilaton and tachyon profile, the coordinate fields for two non-compact directions $X_{L,R}^{7,8}$, as well as their right-moving superpartners $\psi_R^{7,8}$ and left-moving fields $\lambda,\lambda'$ become infinitely massive at late times, decoupling from the worldsheet. The end result is thus a worldsheet CFT describing the background
\begin{equation}\label{chs}
	\mathbb{R}^{5,1}\times\mathbb{R}_\phi \times SU(2)_2 \times (E_8\times E_8)_1,
\end{equation}
where $\mathbb{R}_\phi$ is a linear dilaton direction resulting from 1-loop corrections to the initial lightlike dilaton. This is exactly the CHS background generated by $k = 2$ coincident NS5 branes in the initial theory \cite{Callan:1991at}. More generally, the background of $k$ NS5 branes is described by an $SU(2)/U(1)$ coset model at level $k$, which becomes a trivial CFT precisely at $k = 2$, hence this value for $k$ is not unexpected in this setup.

In spacetime we might interpret this process as producing a local tachyon condensate wherein the non-critical string description applies \cite{Hellerman:2006ff} (see also \cite{Bergman:2006pd}) and the worldvolume of the NS5 brane stack is located along the linear dilaton direction in the strongly coupled region. The change in target space dimensions in the CFT is simply an artifact of describing a localized object as has already been appreciated in the subcritical string literature, see e.g. \cite{Murthy:2006eg}. In this region, gravity decouples in accordance with the mass shift in the graviton due to the linear dilaton. 

\subsubsection{Moduli space analysis}
The starting point of the tachyon condensation process above is the Scherk-Schwarz reduction of the $E_8\times E_8$ string at self-dual radius. The worldsheet sits at a point in the moduli space of non-supersymmetric heterotic strings compactified on $S^1$, and so it corresponds to a Dynkin subdiagram of $\Sigma_{17}$ \cite{Fraiman:2023cpa}:
\begin{eqnarray}
	\begin{tikzpicture}[scale = 1.25]
		\begin{scope}[shift={(-4,0)}]
			\draw(0,0)--(0.5,0.5);
			\draw(2,0)--(1.5,0.5);
			\draw(0,2)--(0.5,1.5);
			\draw(2,2)--(1.5,1.5);
			
			\draw[double, Red](0.5,0.5)--(1,0.75);
			\draw[double, Red](0.5,1.5)--(1,1.25);
			\draw[double, Red](1,0.75)--(1.5,1.5);
			\draw[double, Red, fill=white](1,1.25)--(1.5,0.5);
			\draw[double, Red](1,0.75)--(1,1.25);
			
			\draw(0,0)--(2,0)--(2,2)--(0,2)--(0,0);
			\draw[fill=white](0,0)circle(0.07);
			\draw[fill=white](0.5,0)circle(0.07)node{\textcolor{red}{\ding{55}}};
			\draw[fill=white](1,0)circle(0.07);
			\draw[fill=white](1.5,0)circle(0.07);
			\draw[fill=white](2,0)circle(0.07);
			\draw[fill=white](0,0.5)circle(0.07);
			\draw[fill=white](0,1)circle(0.07);
			\draw[fill=white](0,1.5)circle(0.07);
			\draw[fill=white](2,0.5)circle(0.07);
			\draw[fill=white](2,1)circle(0.07);
			\draw[fill=white](2,1.5)circle(0.07);	
			\draw[fill=white](0,2)circle(0.07);
			\draw[fill=white](0.5,2)circle(0.07);
			\draw[fill=white](1,2)circle(0.07);
			\draw[fill=white](1.5,2)circle(0.07)node{\textcolor{red}{\ding{55}}};
			\draw[fill=white](2,2)circle(0.07);
			
			\draw[fill=white](0.5,0.5)circle(0.07);
			\draw[fill=white](0.5,1.5)circle(0.07);
			\draw[fill=white](1.5,0.5)circle(0.07);
			\draw[fill=white](1.5,1.5)circle(0.07);
			\draw[Red, fill=white,double](1,1.25)circle(0.07)node{\textcolor{red}{\ding{55}}};
			\draw[Red, fill=white,double](1,0.75)circle(0.07);
			\draw[-stealth](2.5,1)--(3.5,1);
			\draw(1,-0.5)node{$\Sigma_{17}$};
		\end{scope}
		
		\draw(0,2)--(0.5,1.5);
		\draw(2,0)--(1.5,0.5);
		
		\draw(1,2)--(0,2)--(0,0);
		\draw(2,2)--(2,0)--(1,0);
		\draw[fill=white](0,0)circle(0.07);
		\draw[fill=white](0.5,2)circle(0.07);
		\draw[fill=white](1,0)circle(0.07);
		%\draw[fill=white](1.5,0)circle(0.07);
		\draw[fill=white](2,0)circle(0.07);
		\draw[fill=white](0,0.5)circle(0.07);
		\draw[fill=white](0,1)circle(0.07);
		\draw[fill=white](0,1.5)circle(0.07);
		\draw[fill=white](2,0.5)circle(0.07);
		\draw[fill=white](2,1)circle(0.07);
		\draw[fill=white](2,1.5)circle(0.07);	
		\draw[fill=white](0,2)circle(0.07);
		%\draw[fill=white](0.5,2)circle(0.07);
		\draw[fill=white](1,2)circle(0.07);
		\draw[fill=white](1.5,0)circle(0.07);
		\draw[fill=white](2,2)circle(0.07);
		
		\draw[fill=white](0.5,1.5)circle(0.07);
		\draw[fill=white](1.5,0.5)circle(0.07);
		\draw[Red, fill=white,double](1,0.75)circle(0.07);
		\draw(1,-0.5)node{$E_8\oplus E_8 \oplus B_1$};
	\end{tikzpicture}
\end{eqnarray}
After the tachyons condense, the red node is deleted and the worldsheet sits at a point in the moduli space described by $\Sigma_{16}$:
\begin{eqnarray}
	\begin{tikzpicture}[scale = 1.25]
		\draw(1,-0.5)node{$\Sigma_{16}$};
		\draw[double, Red](0,0)--(0.75,0.75);
		\draw(2,0)--(1.5,0.5);
		\draw(0,2)--(0.5,1.5);
		\draw[double, Red](2,2)--(1.25,1.25);
		\draw[double, Red](0.75,0.75)--(1.25,1.25);
		
		\draw(0,0)--(2,0)--(2,2)--(0,2)--(0,0);
		\draw[fill=white](0,0)circle(0.07);
		\draw[fill=white](0.5,0)circle(0.07)node{\textcolor{red}{\ding{55}}};
		\draw[fill=white](1,0)circle(0.07);
		\draw[fill=white](1.5,0)circle(0.07);
		\draw[fill=white](2,0)circle(0.07);
		\draw[fill=white](0,0.5)circle(0.07);
		\draw[fill=white](0,1)circle(0.07);
		\draw[fill=white](0,1.5)circle(0.07);
		\draw[fill=white](2,0.5)circle(0.07);
		\draw[fill=white](2,1)circle(0.07);
		\draw[fill=white](2,1.5)circle(0.07);	
		\draw[fill=white](0,2)circle(0.07);
		\draw[fill=white](0.5,2)circle(0.07);
		\draw[fill=white](1,2)circle(0.07);
		\draw[fill=white](1.5,2)circle(0.07)node{\textcolor{red}{\ding{55}}};
		\draw[fill=white](2,2)circle(0.07);
		
		\draw[fill=white](0.5,1.5)circle(0.07);
		\draw[fill=white](1.5,0.5)circle(0.07);
		\draw[Red, fill=white,double](1.25,1.25)circle(0.07)node{\textcolor{red}{\ding{55}}};
		\draw[Red, fill=white,double](0.75,0.75)circle(0.07)node{\textcolor{red}{\ding{55}}};
		\draw[-stealth](2.5,1)--(3.5,1);
		\begin{scope}[shift={(4,0)}]
			\draw(0,2)--(0.5,1.5);
			\draw(2,0)--(1.5,0.5);
			
			\draw(1,2)--(0,2)--(0,0);
			\draw(2,2)--(2,0)--(1,0);
			\draw[fill=white](0,0)circle(0.07);
			\draw[fill=white](0.5,2)circle(0.07);
			\draw[fill=white](1,0)circle(0.07);
			%\draw[fill=white](1.5,0)circle(0.07);
			\draw[fill=white](2,0)circle(0.07);
			\draw[fill=white](0,0.5)circle(0.07);
			\draw[fill=white](0,1)circle(0.07);
			\draw[fill=white](0,1.5)circle(0.07);
			\draw[fill=white](2,0.5)circle(0.07);
			\draw[fill=white](2,1)circle(0.07);
			\draw[fill=white](2,1.5)circle(0.07);	
			\draw[fill=white](0,2)circle(0.07);
			%\draw[fill=white](0.5,2)circle(0.07);
			\draw[fill=white](1,2)circle(0.07);
			\draw[fill=white](1.5,0)circle(0.07);
			\draw[fill=white](2,2)circle(0.07);
			
			\draw[fill=white](0.5,1.5)circle(0.07);
			\draw[fill=white](1.5,0.5)circle(0.07);

			\draw(1,-0.5)node{$E_8\oplus E_8$};
		\end{scope}
	\end{tikzpicture}
\end{eqnarray}

A direct implication of this analysis is that, since the CHS background is supersymmetric, the 1-loop partition function \eqref{ZDS1} with $D = 8, n = 15$ must vanish at the maximal enhancement with gauge group $E_8\times E_8$. This enhancement can be reached from the circle compactification of the $Spin(30)$ sub-critical string by performing an $O(1,16)$ lattice boost 
\begin{equation}
	\Gamma_{1,16}^v \simeq \Gamma_{1,1}\oplus D_{15} \to D_1(-1)\oplus E_8\oplus E_8\,.
\end{equation}
The characters $\mathcal{Z}_{\omega,n}$ all share a sum over the lattice $E_8\oplus E_8$, differing in conjugacy classes of $Spin(2)$. Up to the common $E_8\oplus E_8$ contribution, the terms in parentheses in \eqref{ZDS1} read
\begin{equation}
	\bar O_2 \bar V_6 + \bar V_2 \bar O_6 - \bar C_2 \bar S_6 - \bar S_2 \bar C_6 \simeq \bar V_8 - \bar S_8 = 0\,,
\end{equation}
and $\mathcal{Z}_{S^1}^{8}(\tau,\bar \tau)$ indeed vanishes.\\ 

The procedure carried out for the $E_8\times E_8$ string can be repeated for any heterotic string with at least three noncompact dimensions. For the eight theories in ten dimensions with rank 16, the Scherk-Schwarz reductions all live in the same moduli space. Hence, the CHS background describing a pair of coincident NS5 branes in all of these theories are connected through marginal deformations. Changing from one theory to another is done through T-duality. For the $Spin(32)/\mathbb{Z}_2$ heterotic string the analysis is  straightforward. Simply boost
\begin{equation}
	D_1(-1)\oplus E_8 \oplus E_8 \to D_1(-1) \oplus D_{16}^+\,,
\end{equation}
with $D_{16}^+$ the weight lattice of $Spin(32)/\mathbb{Z}_2$. The other cases are technically more cumbersome to work out, but it should be clear that the statement above holds.

\subsubsection{Duality with 6-branes}\label{sss:duality6}
We have seen that the CHS background \eqref{chs} can be reached by compactifying an eight-dimensional subcritical heterotic string on $S^1$ and suitably tuning the radius and Wilson line moduli. This is particularly true for the $SU(16)/\mathbb{Z}_2$ string (cf. Section \ref{ss:inf}), which has been recently proposed to describe the linear dilaton background generated by a non-supersymmetric 6-brane in the $Spin(32)/\mathbb{Z}_2$ heterotic string \cite{Kaidi:2023tqo}. This brane is charged under $\pi_1 \simeq \mathbb{Z}_2$ of the gauge bundle, and its magnetic flux through $S^2$ breaks the gauge symmetry to $(SU(16)/\mathbb{Z}_2) \times U(1)$; the connection of the tangent bundle of the $S^2$ is then embedded into the $U(1)$ factor. This situation suggests that the 6-brane compactified on $S^1$ with suitable radius and Wilson line becomes supersymmetric and is T-dual to a pair of coincident NS5-branes.   

This T-duality can be seen at the level of the full near horizon background of the brane. The pair of NS5-branes is associated to a 3-sphere $S^3$ threaded by two units of $H_3$ flux, while the 6-brane is associated to a 2-sphere $S^2$ threaded by a gauge flux $F_2 = (\tfrac12^{16})$ with appropriate normalization and a circle $S^1$ with some Wilson line. There must  then exist a topology changing T-duality connecting both backgrounds, which is made possible by the  fact that T-duality in heterotic strings may involve the gauge bundle in non-trivial ways. 

This problem was studied recently in the context of fibrations of $T^2$ over K3 surfaces in \cite{Israel:2023tjw}, and specializing to fibrations of $S^1$ over $S^2$ is straightforward. The heterotic background may be specified by a magnetic charge vector
\begin{equation}
	v = \ket{k,\ell;\lambda}
\end{equation}  
where $k \in \mathbb{Z}$ relates to the first Chern class of the fibration as $k = c_1 \omega$ with $\omega$ the top form of $S^2$, $\ell\in \mathbb{Z}$ similarly encodes the $H_3$-flux and $\lambda \in D_{16}^+$ the $F_2$-flux. The radius $R$ of the fiber and its associated Wilson line moduli $A$ are constrained by flux quantization to satisfy
\begin{equation}\label{fq}
	R^2k + \frac{1}{2}A^2k + A \cdot \lambda= \ell\,.
\end{equation} 
T-duality acts on the background by transforming
\begin{equation}
	v \to v' = Mv\,, ~~~~~ 
\end{equation}
where $M$ is an automorphism of Narain lattice $\text{II}_{1,17}$, and the moduli $R$ and $A$ transform so as to leave equation \eqref{fq} invariant. An important fact is that any positive definite primitive vector $v$ of a given norm is unique up to automorphism in $\text{II}_{1,17}$,\footnote{This is equivalent to stating that any rank 1 (signature $(0,1)$) lattice has a unique (up to automorphisms) primitive embedding into $\Gamma_{1,17}$, which follows from theorem 1.14.4 of \cite{Nikulin}.} meaning that any two such vectors are related by T-duality. The background for a pair of coincident NS5-branes is given by $v = \ket{1,2;0}$, $A = 0$ and $R^2 = 2$. One can T-dualize $v$ to $v' = \ket{0,0;\tfrac12^{16}}$, since $v'^2 = v^2 = 4$, and the resulting background is a direct product $S^1 \times S^2$ with $F_2$-flux threading $S^2$ as desired, and we see that the T-duality picture is consistent.

An interesting consequence of this analysis is that the NS5 background admits 16 marginal deformations which break supersymmetry. It was in fact determined already in \cite{Israel:2004vv} that the sigma model for a 3-sphere $S^3$ admits exactly marginal deformations involving gauge fields such that the $S^3$ is deformed into $S^2 \times S^1$ at infinite distance, where the $S^1$ shrinks to zero size and is T-dualized to $\mathbb{R}$. At this limit, the $H_3$ flux through $S^3$ disappears, and one ends with an $F_2$ flux through the final $S^2$. This matches nicely with our results. What we have done is essentially to clarify that the endpoints of this process for $k = 2$ correspond to subcritical strings with two condensed tachyons, one of which is tachyon-free. With this knowledge in hand, the interpolation found in \cite{Israel:2004vv} can be taken as a proof of the T-duality under consideration. We see that the marginal deformations found in that work for $k = 2$ span precisely the moduli space $O(1,16;\mathbb{Z})\backslash O(1,16)/O(16)$, and it would be interesting to generalize this result to other values of $k$.

\subsection{Intersections of branes}
\subsubsection{An NS5 pair and a 6-brane}
Consider now the Scherk-Schwarz reduction of the non-supersymmetric $U(16)$ heterotic string at self-dual radius. This configuration has two pairs of tachyons, each of which condenses respectively to the linear dilaton background of a 6-brane compactified on $S^1$ and an NS5-brane pair, hence after condensing both pairs we expect to have a model associated to the intersection of these objects. From our discussion above, the resulting background reads
\begin{equation}\label{ns56}
	\mathbb{R}^{3,1}\times \mathbb{R}_\phi \times SU(2)_2 \times (SU(16)/\mathbb{Z}_2)_1\,.
\end{equation}

Since we are condensing four tachyons, the end result sits at a point in the moduli space described by the diagram $\Sigma_{15}$:
\begin{eqnarray}
	\begin{tikzpicture}[scale = 1.25]
	\draw(1,-0.5)node{$\Sigma_{15}$};
	\draw(2,0)--(1.5,0.5);
	\draw(0,2)--(0.5,1.5);
	
	\draw(0,0)--(2,0)--(2,2)--(0,2)--(0,0);
	
	\draw[Red, fill=white,double](1.5,2)--(2,2);
	\draw[Red, fill=white,double](2,1.5)--(2,2);
	
	\draw[fill=white](0,0)circle(0.07);
	\draw[fill=white](0.5,0)circle(0.07);
	\draw[fill=white](1,0)circle(0.07);
	\draw[fill=white](1.5,0)circle(0.07);
	\draw[fill=white](2,0)circle(0.07);
	\draw[fill=white](0,0.5)circle(0.07);
	\draw[fill=white](0,1)circle(0.07);
	\draw[fill=white](0,1.5)circle(0.07);
	\draw[fill=white](2,0.5)circle(0.07);
	\draw[fill=white](2,1)circle(0.07);
	\draw[fill=white](2,1.5)circle(0.07);	
	\draw[fill=white](0,2)circle(0.07);
	\draw[fill=white](0.5,2)circle(0.07);
	\draw[fill=white](1,2)circle(0.07);
	\draw[fill=white](1.5,2)circle(0.07);
	
	\draw[fill=white](0.5,1.5)circle(0.07)node{\textcolor{red}{\ding{55}}};
	\draw[fill=white](1.5,0.5)circle(0.07)node{\textcolor{red}{\ding{55}}};
	\draw[Red, fill=white,double](2,2)circle(0.07)node{\textcolor{red}{\ding{55}}};
		\draw[-stealth](2.5,1)--(3.5,1);
		\begin{scope}[shift={(4,0)}]
			\draw(1,-0.5)node{$A_{15}$};

			\draw(0,0)--(2,0)--(2,1.5);
			\draw(0,0)--(0,2)--(1.5,2);
			
			\draw[fill=white](0,0)circle(0.07);
			\draw[fill=white](0.5,0)circle(0.07);
			\draw[fill=white](1,0)circle(0.07);
			\draw[fill=white](1.5,0)circle(0.07);
			\draw[fill=white](2,0)circle(0.07);
			\draw[fill=white](0,0.5)circle(0.07);
			\draw[fill=white](0,1)circle(0.07);
			\draw[fill=white](0,1.5)circle(0.07);
			\draw[fill=white](2,0.5)circle(0.07);
			\draw[fill=white](2,1)circle(0.07);
			\draw[fill=white](2,1.5)circle(0.07);	
			\draw[fill=white](0,2)circle(0.07);
			\draw[fill=white](0.5,2)circle(0.07);
			\draw[fill=white](1,2)circle(0.07);
			\draw[fill=white](1.5,2)circle(0.07);
		\end{scope}
	\end{tikzpicture}
\end{eqnarray}
The partition function is given by eq. \eqref{ZDS1} with $D = 6$ and $n = 14$, with boosted momentum lattice
\begin{equation}
	\Gamma_{1,1}\oplus D_{14} \to [D_1(-1)\oplus A_{15}^{(8)}]^{(2,4)}\,,
\end{equation}
where $A_{15}^{(8)}$ is the weight lattice of $SU(16)/\mathbb{Z}_2$ and $(2,4)$ is a gluing vector $y = y_1+y_2$ with $y_{1,2}$ index 2 vectors in $D_1(-1)$ and $A_{15}^{(8)}$. 

It would be interesting to study the stability of this configuration. It would not be unreasonable to expect it to be stable given that the moduli fields are associated to deformations of a supersymmetric background. Supersymmetry is broken by the presence of the 6-brane, which has no such moduli. Also note that this picture generalizes easily to intersections of NS5 pairs with other non-supersymmetric branes by starting with the appropriate subcritical string.

Using the duality between the NS5 pair and the 6-brane, we may interpret the background \eqref{ns56} as being produced by an intersection of two 6-branes compactified on a circle with suitable radius and Wilson line. Similarly, we may consider this background as resulting from a circle compactification of the $E_7\times E_7$ subcritical heterotic string, suggesting a duality with the heterotic 4-brane of \cite{Kaidi:2023tqo}. Indeed, an intersection of two 6-branes has 4 space dimensions. This, however, poses a problem. The background of the 4-brane consists in a 4-sphere $S^4$ with (anti)instantons breaking each $E_8$ to $E_7$ in the $E_8\times E_8$ heterotic string. For consistency we would require a topology changing T-duality transforming this $S^4$ into a product $S^2\times S^2$, with each $S^2$ threaded respectively by an appropriate gauge flux. Such a possibility seems unlikely, suggesting that many naive duality relationships between brane configurations could be an artifact of looking only at the theories at low energies where the sphere contributions are gapped. 

\subsubsection{Two NS5 pairs}\label{ss:intns5}
Consider now two consecutive Scherk-Schwarz reductions performed on some critical heterotic string, say the $E_8 \times E_8$ string. Take the associated circles to be perpendicular and both at self-dual radius. The resulting background has two pairs of tachyons, each of which may be condensed just as for the case of one Scherk-Schwarz reduction explained in Section \ref{ss:ns5s}. The resulting background reads
\begin{equation}\label{intns5s}
	\mathbb{R}^{2,1}\times \mathbb{R}_\phi \times SU(2)_2 \times SU(2)_2 \times (E_8 \times E_8)_1\,.
\end{equation}
This background corresponds to two intersecting pairs of coincident NS5-branes. 

The moduli space for the background \eqref{intns5s} is more complicated than those we have considered mainly in this paper, since it is associated a $T^2$ compactification of a subcritical string. Still it is straightforward to generalize some results from the $S^1$ case. Namely the vector and scalar spacetime classes furnish the odd self-dual lattice $\text{I}_{2,16}$, and the T-duality group reads $O(2,16;\mathbb{Z})$. An analysis of this group is far outside the scope of this paper, but we may still analyze the moduli space using elementary lattice manipulations. 

The vector class lattice $\Gamma_v$ takes the form $\Gamma_{2,2}\oplus D_{14}$, which at the specific point \eqref{intns5s} is polarized to $D_{2}(-1)\oplus E_8 \oplus E_8$. Equivalently, one may write
\begin{equation}
	\Gamma_v \simeq \Gamma_{v,1}\oplus \Gamma_{v,2}\,, ~~~~~ \Gamma_{v,1} \simeq \Gamma_{v,2} \simeq D_1 \oplus E_8\,,
\end{equation}
and separately polarize each of the two sublattices in the sum. These can be polarized in particular to $\Gamma_{1,1}\oplus E_7$, making manifest the realization of the background as a torus compactification of the subcritical $E_7 \times E_7$ heterotic string.
As before, this would naively suggest that two intersecting pairs of NS5 branes are T-dual to a torus compactification of the non-supersymmetric 4-brane with suitable values for the moduli. At the level of the $S^3\times S^3$ sigma model, however, a topology changing T-duality simply leads to $S^2\times S^2 \times S^1\times S^1$ with appropriate gauge flux threading both 2-spheres, i.e. an intersection of two 6-branes compactified on $S^1 \times S^1$. It would be interesting to understand this relationship further.

\section{Conclusions}
\label{s:conc}

In this note we have determined the global structure of the moduli (deformation) spaces of non-critical heterotic strings with maximal rank compactified on a circle, explicitly up to 14 spacetime dimensions. The T-duality groups turned out to be described by Coxeter polytopes following the work of Vinberg \cite{Vinberg:1972} and Vinverg-Kaplinskaya \cite{Kaplinskaya:1978}. We have then used these results to systematically extend the application of ``dimension changing" tachyon condensation of Hellerman-Swanson \cite{Hellerman:2006ff} from tachyonic non-supersymmetric heterotic strings \cite{Kaidi:2020jla} to their circle compactifications in the case of maximal rank (i.e. omitting the $E_8$ string). 

Inspired by the connection between these backgrounds and certain non-supersymmetric heterotic branes \cite{Kaidi:2023tqo} we have extended this interpretation to various other backgrounds, establishing a connection between the momentum-winding tachyons of the heterotic string on a Scherk-Schwarz circle with a pair of NS5 branes, providing a possible tachyon condensation mechanism for these objects in a spirit similar to earlier results which relied on the use of supercritical strings \cite{Garcia-Etxebarria:2014txa}. Along the way we have pointed out the possibility that a pair of coincident NS5 branes is dual to the non-supersymmetric 6-brane of \cite{Kaidi:2023tqo} compactified on a circle with special radius and Wilson lines.

An important caveat that we have not addressed in this paper concerns the regularization of the strong coupling region in the linear dilaton backgrounds for subcritical strings.\footnote{We thank S. Sethi for emphasizing this point to us.} For the supersymmetric configurations corresponding to NS5-branes, this may be done by trading the $\mathbb{R}_\phi \times \mathbb{R}_t$ part of the CFT by the coset $SL(2,\mathbb{R})_2/U(1)$, see e.g. \cite{Giveon:2006pr,Murthy:2006eg}. Without this regularization it is not clear a priori what physical information can be extracted by computing the 1-loop partition function of the compactified models. We expect that the resulting behavior is analogous to that of the critical case studied in \cite{Fraiman:2023cpa}, up to novelties such as supersymmetry enhancement, which is possible due to the absence of gravity in these models. 

It would be of great interest to better understand the spacetime picture of brane production through tachyon condensation. This seems to be the right interpretation of the setups of Hellerman-Swanson-Kaidi \cite{Hellerman:2006ff,Kaidi:2020jla}, rather than resulting in a quantum gravity vacuum with positive cosmological constant -- indeed, Lorentz symmetry is broken in these setups and more importantly the graviton degrees of freedom are massed up, falling outside of the class of constructions that are relevant for the study of generic properties of quantum gravity theories, i.e. the Swampland program \cite{Agmon:2022thq}. Not all is lost, however, as this mechanism might help in understanding possible spacetime deffects in heterotic strings such as the branes of \cite{Kaidi:2023tqo}, informing us how the cobordism conjecture \cite{McNamara:2019rup} applies in them (see also \cite{Basile:2023knk}).

Finally, it has been recently conjectured \cite{Yonekura:2024spl} that an NS $p$-brane compactified on $S^{8-q}$ is dual to a $q$-brane on $S^{8-p}$ with appropriate fluxes on the spheres. One such setup corresponds to what we have interpreted here as the background of an intersection of the non-supersymmetric heterotic 6-brane and a stack of $n$ coincident NS5 branes, where we have specialized to $n = 2$. In \cite{Yonekura:2024spl} the aforementioned duality was used to make inferences on the worldvolume content of the 6-brane. It would be interesting to connect this picture with that of the duality between the 6-brane on $S^1$ and the $n = 2$ NS5 brane stack.

\section*{Acknowledgements}

We thank Soumangsu Chakraborty, Bernardo Fraiman, Ruben Minasian, Savdeep Sethi and Cumrun Vafa for useful discussions. We thank Miguel Montero for very helpful correspondence, and also the IPhT, Saclay for hospitality. This work is supported by a grant from the Simons Foundation (602883,CV), the DellaPietra Foundation and by the NSF grant PHY2013858.

\appendix

\bibliographystyle{JHEP}
\bibliography{branes}

\providecommand{\href}[2]{#2}\begingroup\raggedright\begin{thebibliography}{10}

\bibitem{Blumenhagen:2013fgp}
R.~Blumenhagen, D.~L\"ust, and S.~Theisen, {\em {Basic concepts of string
  theory}}.
\newblock Theoretical and Mathematical Physics. Springer, Heidelberg, Germany,
  2013.

\bibitem{Narain:1985jj}
K.~S. Narain, {\it {New Heterotic String Theories in Uncompactified Dimensions
  \ensuremath{<} 10}},  {\em Phys. Lett. B} {\bf 169} (1986) 41--46.

\bibitem{Narain:1986am}
K.~S. Narain, M.~H. Sarmadi, and E.~Witten, {\it {A Note on Toroidal
  Compactification of Heterotic String Theory}},  {\em Nucl. Phys. B} {\bf 279}
  (1987) 369--379.

\bibitem{Cachazo:2000ey}
F.~A. Cachazo and C.~Vafa, {\it {Type I' and real algebraic geometry}},
  \href{http://arxiv.org/abs/hep-th/0001029}{{\tt hep-th/0001029}}.

\bibitem{Fraiman:2018ebo}
B.~Fraiman, M.~Gra\~na, and C.~A. N\'u\~nez, {\it {A new twist on heterotic
  string compactifications}},  {\em JHEP} {\bf 09} (2018) 078,
  [\href{http://arxiv.org/abs/1805.11128}{{\tt arXiv:1805.11128}}].

\bibitem{Font:2020rsk}
A.~Font, B.~Fraiman, M.~Gra\~na, C.~A. N\'u\~nez, and H.~P. De~Freitas, {\it
  {Exploring the landscape of heterotic strings on $T^d$}},  {\em JHEP} {\bf
  10} (2020) 194, [\href{http://arxiv.org/abs/2007.10358}{{\tt
  arXiv:2007.10358}}].

\bibitem{Vinberg:1972}
E.~B. Vinberg, {\it On groups of unit elements of certain quadratic forms},
  {\em Math. USSR, Sb.} {\bf 16} (1972) 17--35.

\bibitem{Chamseddine:1991qu}
A.~H. Chamseddine, {\it {A Study of noncritical strings in arbitrary
  dimensions}},  {\em Nucl. Phys. B} {\bf 368} (1992) 98--120.

\bibitem{Hellerman:2004zm}
S.~Hellerman, {\it {On the landscape of superstring theory in D \ensuremath{>}
  10}},  \href{http://arxiv.org/abs/hep-th/0405041}{{\tt hep-th/0405041}}.

\bibitem{Murthy:2006eg}
S.~Murthy, {\it {Non-critical heterotic superstrings in various dimensions}},
  {\em JHEP} {\bf 10} (2006) 037,
  [\href{http://arxiv.org/abs/hep-th/0603121}{{\tt hep-th/0603121}}].

\bibitem{Kaplinskaya:1978}
{\`E}.~B. Vinberg and I.~M. Kaplinskaya, {\it On the groups
  {{\(O_{18,1}(\mathbb Z)\)}} and {{\(O_{19,1}(\mathbb Z)\)}}},  {\em Sov.
  Math., Dokl.} {\bf 19} (1978) 194--197.

\bibitem{Seiberg:2005nk}
N.~Seiberg, {\it {Long strings, anomaly cancellation, phase transitions,
  T-duality and locality in the 2-D heterotic string}},  {\em JHEP} {\bf 01}
  (2006) 057, [\href{http://arxiv.org/abs/hep-th/0511220}{{\tt
  hep-th/0511220}}].

\bibitem{Davis:2005qe}
J.~L. Davis, F.~Larsen, and N.~Seiberg, {\it {Heterotic strings in two
  dimensions and new stringy phase transitions}},  {\em JHEP} {\bf 08} (2005)
  035, [\href{http://arxiv.org/abs/hep-th/0505081}{{\tt hep-th/0505081}}].

\bibitem{Davis:2005qi}
J.~L. Davis, {\it {The Moduli space and phase structure of heterotic strings in
  two dimensions}},  {\em Phys. Rev. D} {\bf 74} (2006) 026004,
  [\href{http://arxiv.org/abs/hep-th/0511298}{{\tt hep-th/0511298}}].

\bibitem{Kaidi:2023tqo}
J.~Kaidi, K.~Ohmori, Y.~Tachikawa, and K.~Yonekura, {\it {Nonsupersymmetric
  Heterotic Branes}},  {\em Phys. Rev. Lett.} {\bf 131} (2023), no.~12 121601,
  [\href{http://arxiv.org/abs/2303.17623}{{\tt arXiv:2303.17623}}].

\bibitem{Polchinski:2005bg}
J.~Polchinski, {\it {Open heterotic strings}},  {\em JHEP} {\bf 09} (2006) 082,
  [\href{http://arxiv.org/abs/hep-th/0510033}{{\tt hep-th/0510033}}].

\bibitem{Bergshoeff:2006bs}
E.~A. Bergshoeff, G.~W. Gibbons, and P.~K. Townsend, {\it {Open M5-branes}},
  {\em Phys. Rev. Lett.} {\bf 97} (2006) 231601,
  [\href{http://arxiv.org/abs/hep-th/0607193}{{\tt hep-th/0607193}}].

\bibitem{Kaidi:2020jla}
J.~Kaidi, {\it {Stable Vacua for Tachyonic Strings}},  {\em Phys. Rev. D} {\bf
  103} (2021), no.~10 106026, [\href{http://arxiv.org/abs/2010.10521}{{\tt
  arXiv:2010.10521}}].

\bibitem{Callan:1991at}
C.~G. Callan, Jr., J.~A. Harvey, and A.~Strominger, {\it {Supersymmetric string
  solitons}},  \href{http://arxiv.org/abs/hep-th/9112030}{{\tt
  hep-th/9112030}}.

\bibitem{Baykara:2024tjr}
Z.~K. Baykara, H.-C. Tarazi, and C.~Vafa, {\it {New Non-Supersymmetric
  Tachyon-Free Strings}},  \href{http://arxiv.org/abs/2406.00185}{{\tt
  arXiv:2406.00185}}.

\bibitem{Angelantonj:2024jtu}
C.~Angelantonj, I.~Florakis, G.~Leone, and D.~Perugini, {\it
  {Non-supersymmetric non-tachyonic heterotic vacua with reduced rank in
  various dimensions}},  \href{http://arxiv.org/abs/2407.09597}{{\tt
  arXiv:2407.09597}}.

\bibitem{Kawai:1986vd}
H.~Kawai, D.~C. Lewellen, and S.~H.~H. Tye, {\it {Classification of Closed
  Fermionic String Models}},  {\em Phys. Rev. D} {\bf 34} (1986) 3794.

\bibitem{Seiberg:1986by}
N.~Seiberg and E.~Witten, {\it {Spin Structures in String Theory}},  {\em Nucl.
  Phys. B} {\bf 276} (1986) 272.

\bibitem{Berasaluce-Gonzalez:2013sna}
M.~Berasaluce-Gonz\'alez, M.~Montero, A.~Retolaza, and A.~M. Uranga, {\it
  {Discrete gauge symmetries from (closed string) tachyon condensation}},  {\em
  JHEP} {\bf 11} (2013) 144, [\href{http://arxiv.org/abs/1305.6788}{{\tt
  arXiv:1305.6788}}].

\bibitem{Fraiman:2023cpa}
B.~Fraiman, M.~Gra\~na, H.~Parra De~Freitas, and S.~Sethi, {\it
  {Non-Supersymmetric Heterotic Strings on a Circle}},
  \href{http://arxiv.org/abs/2307.13745}{{\tt arXiv:2307.13745}}.

\bibitem{Borcherds:1987}
R.~Borcherds, {\it Automorphism groups of lorentzian lattices},  {\em Journal
  of Algebra} {\bf 111} (1987), no.~1 133--153.

\bibitem{Collazuol:2022jiy}
V.~Collazuol, M.~Gra\~na, and A.~Herr\'aez, {\it {E$_{9}$ symmetry in the
  heterotic string on S$^{1}$ and the weak gravity conjecture}},  {\em JHEP}
  {\bf 06} (2022) 083, [\href{http://arxiv.org/abs/2203.01341}{{\tt
  arXiv:2203.01341}}].

\bibitem{Collazuol:2022oey}
V.~Collazuol, M.~Gra\~na, A.~Herr\'aez, and H.~Parra De~Freitas, {\it {Affine
  algebras at infinite distance limits in the Heterotic String}},  {\em JHEP}
  {\bf 07} (2023) 036, [\href{http://arxiv.org/abs/2210.13471}{{\tt
  arXiv:2210.13471}}].

\bibitem{Collazuol:2024kzl}
V.~Collazuol and I.~V. Melnikov, {\it {A twist at infinite distance in the CHL
  string}},  \href{http://arxiv.org/abs/2402.01606}{{\tt arXiv:2402.01606}}.

\bibitem{Ginsparg:1986wr}
P.~H. Ginsparg and C.~Vafa, {\it {Toroidal Compactification of
  Nonsupersymmetric Heterotic Strings}},  {\em Nucl. Phys. B} {\bf 289} (1987)
  414.

\bibitem{BoyleSmith:2023xkd}
P.~Boyle~Smith, Y.-H. Lin, Y.~Tachikawa, and Y.~Zheng, {\it {Classification of
  chiral fermionic CFTs of central charge $\le$ 16}},  {\em SciPost Phys.} {\bf
  16} (2024), no.~2 058, [\href{http://arxiv.org/abs/2303.16917}{{\tt
  arXiv:2303.16917}}].

\bibitem{Hohn:2023auw}
G.~H\"ohn and S.~M\"oller, {\it {Classification of Self-Dual Vertex Operator
  Superalgebras of Central Charge at Most 24}},
  \href{http://arxiv.org/abs/2303.17190}{{\tt arXiv:2303.17190}}.

\bibitem{Rayhaun:2023pgc}
B.~C. Rayhaun, {\it {Bosonic Rational Conformal Field Theories in Small Genera,
  Chiral Fermionization, and Symmetry/Subalgebra Duality}},
  \href{http://arxiv.org/abs/2303.16921}{{\tt arXiv:2303.16921}}.

\bibitem{King:2003}
O.~D. {King}, {\it {A mass formula for unimodular lattices with no roots}},
  {\em Mathematics of Computation} {\bf 72} (Jan., 2003) 839--863,
  [\href{http://arxiv.org/abs/math/0012231}{{\tt math/0012231}}].

\bibitem{Hellerman:2006ff}
S.~Hellerman and I.~Swanson, {\it {Dimension-changing exact solutions of string
  theory}},  {\em JHEP} {\bf 09} (2007) 096,
  [\href{http://arxiv.org/abs/hep-th/0612051}{{\tt hep-th/0612051}}].

\bibitem{Hellerman:2007zz}
S.~Hellerman and I.~Swanson, {\it {A Stable vacuum of the tachyonic E(8)
  string}},  \href{http://arxiv.org/abs/0710.1628}{{\tt arXiv:0710.1628}}.

\bibitem{Guglielmetti:2015}
R.~Guglielmetti, {\it Coxiter – computing invariants of hyperbolic coxeter
  groups},  {\em LMS Journal of Computation and Mathematics} {\bf 18} (2015),
  no.~1 754–773.

\bibitem{Lerche:1986he}
W.~Lerche and D.~Lust, {\it {Covariant Heterotic Strings and Odd Selfdual
  Lattices}},  {\em Phys. Lett. B} {\bf 187} (1987) 45--50.

\bibitem{Israel:2004vv}
D.~Israel, C.~Kounnas, D.~Orlando, and P.~M. Petropoulos, {\it
  {Electric/magnetic deformations of S**3 and AdS(3), and geometric cosets}},
  {\em Fortsch. Phys.} {\bf 53} (2005) 73--104,
  [\href{http://arxiv.org/abs/hep-th/0405213}{{\tt hep-th/0405213}}].

\bibitem{Atick:1988si}
J.~J. Atick and E.~Witten, {\it {The Hagedorn Transition and the Number of
  Degrees of Freedom of String Theory}},  {\em Nucl. Phys. B} {\bf 310} (1988)
  291--334.

\bibitem{Saxena:2024eil}
V.~Saxena, {\it {A T-Duality of Non-Supersymmetric Heterotic Strings and an
  implication for Topological Modular Forms}},
  \href{http://arxiv.org/abs/2405.19409}{{\tt arXiv:2405.19409}}.

\bibitem{Bergman:2006pd}
O.~Bergman and S.~S. Razamat, {\it {Toy models for closed string tachyon
  solitons}},  {\em JHEP} {\bf 11} (2006) 063,
  [\href{http://arxiv.org/abs/hep-th/0607037}{{\tt hep-th/0607037}}].

\bibitem{Israel:2023tjw}
D.~Israel, I.~V. Melnikov, R.~Minasian, and Y.~Proto, {\it {Topology change and
  heterotic flux vacua}},  \href{http://arxiv.org/abs/2312.08923}{{\tt
  arXiv:2312.08923}}.

\bibitem{Nikulin}
V.~V. Nikulin, {\it Integral symmetric bilinear forms and some of their
  applications},  {\em Mathematics of the USSR-Izvestiya} {\bf 14} (feb, 1980)
  103.

\bibitem{Garcia-Etxebarria:2014txa}
I.~n. Garc\'\i{}a-Etxebarria, M.~Montero, and A.~Uranga, {\it {Heterotic
  NS5-branes from closed string tachyon condensation}},  {\em Phys. Rev. D}
  {\bf 90} (2014), no.~12 126002, [\href{http://arxiv.org/abs/1405.0009}{{\tt
  arXiv:1405.0009}}].

\bibitem{Giveon:2006pr}
A.~Giveon and D.~Kutasov, {\it {Fundamental strings and black holes}},  {\em
  JHEP} {\bf 01} (2007) 071, [\href{http://arxiv.org/abs/hep-th/0611062}{{\tt
  hep-th/0611062}}].

\bibitem{Agmon:2022thq}
N.~B. Agmon, A.~Bedroya, M.~J. Kang, and C.~Vafa, {\it {Lectures on the string
  landscape and the Swampland}},  \href{http://arxiv.org/abs/2212.06187}{{\tt
  arXiv:2212.06187}}.

\bibitem{McNamara:2019rup}
J.~McNamara and C.~Vafa, {\it {Cobordism Classes and the Swampland}},
  \href{http://arxiv.org/abs/1909.10355}{{\tt arXiv:1909.10355}}.

\bibitem{Basile:2023knk}
I.~Basile, A.~Debray, M.~Delgado, and M.~Montero, {\it {Global anomalies \&
  bordism of non-supersymmetric strings}},  {\em JHEP} {\bf 02} (2024) 092,
  [\href{http://arxiv.org/abs/2310.06895}{{\tt arXiv:2310.06895}}].

\bibitem{Yonekura:2024spl}
K.~Yonekura, {\it {Dualities among Neveu-Schwarz sector branes in string
  theory}},  \href{http://arxiv.org/abs/2403.14933}{{\tt arXiv:2403.14933}}.

\end{thebibliography}\endgroup

\end{document}